\documentclass[12pt]{amsart}
\usepackage{amsbsy,amssymb,amsmath,amsthm,amscd,amsfonts,latexsym,amstext,delarray,
amsmath,graphicx} 
\usepackage[margin=1in]{geometry}
\usepackage{color}

\newtheorem{thm}{Theorem}[section]
\newtheorem{prop}{Proposition}[section]
\newtheorem{cor}[thm]{Corollary}
\newtheorem{lem}[thm]{Lemma}

\newtheorem{defn}[prop]{Definition}
\newtheorem{rem}[prop]{Remark}
\newtheorem{ex}[prop]{Example}
\newtheorem{ques}{Question}

\def\N{{\mathbb N}}
\def\R{{\mathbb R}}
\def\C{{\mathbb C}}

\def\cB{{\mathcal B}}
\def\cC{{\mathcal C}}
\def\cD{{\mathcal D}}

\def\cF{{\mathcal F}}
\def\cG{{\mathcal G}}

\def\cI{{\mathcal I}}
\def\cJ{{\mathcal J}}

\def\cL{{\mathcal L}}
\def\cM{{\mathcal M}}
\def\cN{{\mathcal N}}

\def\cP{{\mathcal P}}

\def\cR{{\mathcal R}}
\def\cS{{\mathcal S}}
\def\cT{{\mathcal T}}

\def\cX{{\mathcal X}}

\def\bD{{\mathbb D}}

\def\bL{{\mathbb L}}

\def\bS{{\mathbb S}}

\def\fA{{\mathfrak A}}
\def\fB{{\mathfrak B}}
\def\fC{{\mathfrak C}}
\def\fD{{\mathfrak D}}
\def\fQ{{\mathfrak Q}}
\def\fS{{\mathfrak S}}

\def\Tr{{\rm Tr}}

\title{Formal languages, spin systems, and quasicrystals}
\author{Francesca Fernandes and Matilde Marcolli}
\address{Physics Department, Stanford University,  
Stanford, CA 94305, USA}
\email{fcf@stanford.edu}
\address{Mathematics Department, Caltech, Pasadena, CA 91125, USA}
\email{matilde@caltech.edu}
\date{}

\begin{document}
\maketitle

\begin{abstract}
We present a categorical formalism for context-free languages with morphisms
given by correspondences obtained from rational transductions. We show that
D0L-systems are a special case of the correspondences that define morphisms
in this category. We construct 
a functorial mapping to aperiodic spin chains. We then generalize this construction to a class
of mildly context sensitive grammars, the multiple-context-free grammars (MCFG),
with a similar functorial mapping to spin systems in higher dimensions, with Boltzmann weights 
describing interacting spins on vertices of hypercubes. We show that a particular motivating
example for this general construction is provided by the Korepin completely
integrable model on the icosahedral quasicrystal, which we construct as the
spin system associated to a multiple-context-free grammar describing 
the geometry of the Ammann planes quasilattice. We review the main properties
of this spin system, including solvability, bulk free energy, and criticality, 
based on results of Baxter and the known relation to the Zamolodchikov tetrahedron 
equation, and we show that the latter has a generalization for the Boltzmann
weights on hypercubes of the spin systems associated to more general MCFGs 
in terms of two dual cubulations of the $n$-simplex.
We formulate analogous questions about bulk free energy and criticality 
for our construction of spin systems. 
\end{abstract}

\section{Introduction}

Originally associated to the context of generative linguistics (\cite{Chomsky1}, \cite{Chomsky2}), 
and broadly used in computer science and computational linguistics, the 
mathematical theory of formal languages has also found many significant 
applications to theoretical physics, 
see for instance \cite{DoIsh}, \cite{GuImKaKhoLi}, \cite{LinTeg}, \cite{LinTeg2},
\cite{MaPo}. We focus here on the relation between formal languages
and spin models. 

Spin models study physical systems with spins arranged in a certain formation. 
They enable the study of magnets, quantum computing, and phase transitions 
(see for instance \cite{Micheli}, \cite{Evans}, \cite{Shiina}, \cite{Derkachov}).
In two dimensions, many exactly solvable spin models have been 
constructed \cite{Bax4}, \cite{Bax5}, \cite{Popkov}. Some 
exactly solvable spin models have also been constructed in three dimensions
(see for instance  \cite{BazhBax}, \cite{Bravyi}, \cite{Mandal}). More generally,
there are no explicit algebraic methods for the constructions of classes of spin systems 
in higher dimensions. A goal of this paper is to develop such a method,
based on a categorical formulation of context-free and multiple context-free
formal languages. The construction method that we propose will in general
produce spin systems that need not be exactly solvable models,
though it includes well known cases that are exactly solvable (as we will
show explicitly in the case of the 3D icosahedral quasicrystal) and it
gives rise to a natural higher dimensional generalization of the tetrahedron
equation for the Boltzmann weights, expressed in terms of two dual cubulations
of the $n$-simplex.  A main feature that the models we describe have in 
general is a general treatment of
substitution rules, formulated in terms of transductions in the underlying 
formal languages. These include as a special case
the inflation rules of D0L-systems describing aperiodic spin chains. 

The relation between formal languages and spin models is well known 
in the treatment of aperiodic spin chains, see  \cite{BaGri}. Moreover,
a general result of \cite{StDrDeC} shows that, while classical spin Hamiltonians
in dimension one are modeled by context-free formal languages, in dimension
two and higher they require context-sensitive grammars. The analysis of
\cite{StDrDeC} focuses on the case of all-to-all spin Hamiltonians. Our viewpoint
here is different, in the sense that we use formal languages to describe
aperiodic structures and inflation/decimation rules for spin systems, but in
this setting we also find that a class of mildly context-sensitive grammars,
the multiple context-free grammars (MCFG) can be used to describe a class of 
vertex-model-type spin systems that include the Zamolodchikov model in 3D and
Korepin's completely solvable spin model on the icosahedral quasicrystal.

We present here a new mathematical 
perspective, based on categorical properties and functoriality, and a
natural way, in this setting, of passing from the class of context-free formal 
languages to the mildly context-sensitive class of multiple context-free grammars. 
In general, the spin systems that we will obtain in this way do not correspond to
actual geometries: they are abstract models where the encoding by formal
languages resembles the properties of quasiperiodic systems in actual
geometric cases. The class of systems we construct does, however, include
some specific geometric cases. 
We describe the specific example of the Korepin integrable spin system on the 
3D icosahedral quasicrystal as an explicit illustration and as
the main motivating example for our general construction. 

Boyle and Steinhardt \cite{BoStein} developed a construction of
quasicrystals based on Coxeter pairs describing a superposition
structure of Ammann patterns, which are sets of 1D quasi-periodic 
arrangements of hyperplanes. They classify the Penrose-like tiling dual to these 
Ammann patterns: in the 2D case there are $11$ such Penrose tilings, 
$9$ in the 3D case, and a single 4D case, with none in higher dimensions. 
We note here, by focusing on a 3D case as
a main example, that this Ammann pattern construction is 
especially suitable for description in terms of formal languages. 
In \S \ref{AmmannFLsec} give a construction of the Korepin exactly solvable 
spin model for the icosahedral quasicrystal introduced in \cite{Ko},
based on a multiple-context-free grammar encoding the geometry of
the Ammann patterns. 
In two dimensions, exactly solvable spin models satisfy the star-triangle 
equations \cite{Bax}. In three dimensions, these two-dimensional star-triangle 
equations generalize to the three-dimensional Zamolodchikov tetrahedron equation \cite{Zamo},
and Baxter's solution in \cite{Bax2}. We show that the tetrahedron equation in terms
of the Boltzmann weights as in \cite{Bax2} has a natural extension to the more
general spin systems from multiple-context-free grammars, in the form of an
$n$-simplex equation. We also discuss the bulk free energy and the criticality property
of the Korepin spin model on the icosahedral quasicrystal and formulate related
questions for our more general spin systems. 

\subsection{Formal languages and aperiodic structures}

Quasicrystals have fascinated scientists and mathematicians alike since their discovery in 1987. 
Unlike traditional crystals, they lack translational symmetry, 
and only possess rotational symmetries of orders five or greater than six.
They are found in composite alloys and possess intriguing properties 
such as superior hardness, low thermal conductivity, and an ability to 
absorb hydrogen \cite{Janot}. 
One dimensional aperiodic spin chains are widely studied, see \cite{BaGri}.
In two dimensions, quasicrystals are mathematically represented using aperiodic 
tilings, and in three dimensions, using aperiodic packings, see \cite{BaGri}, \cite{Janot}, \cite{Sen}.
A very well studied quasicrystal is 
the 3D icosahedral quasicrystal, which can be seen as a generalization
of the two-dimensional Penrose tiling and are based on  
three-dimensional icosahedral packings,  \cite{Mad}.
Packings corresponding to a three-dimensional quasicrystal can be 
very complicated. Thus, representing these quasicrystals mathematically 
can be challenging. We argue that 
formal languages can ameliorate this issue, providing a powerful 
yet compact way to represent quasicrystal structures.
Yet despite their utility, only one formal language has so far being
considered for a three-dimensional quasicrystal \cite{GarEsc1}, \cite{GarEsc2}. 
In \S \ref{IcosaqlattSec} we present as one of our
main motivating examples, a completely integrable spin model 
on the icosahedral quasicrystal.  This is one of few exactly solvable 
three-dimensional spin models, and we argue that the formal languages
perspective enables further computational study of this spin model and 
of the geometric and physical properties of the icosahedral quasicrystal. 
We use the dual Ammann quasilattice to construct a formal language 
for the icosahedral quasicrystal and we then use the formal language to
describe the exactly solvable spin model on the icosahedral quasicrystal.
We compute the bulk free energy following \cite{Ko} and we show that
criticality holds as expected in \cite{Bax}.

\section{Formal languages: a categorical viewpoint}\label{FLcatSec}

We focus here on the context-free class of formal languages and
we present it in a categorical framework. This will be helpful in 
the next section, where we describe our construction of associated 
spin models, which will then have nice functorial properties. Our
categorical construction here is based on using certain {\em rational transducers} 
as a good class of morphisms. We refer the reader to \cite{Berstel} for
general background on transducers.

\smallskip
\subsection{Context-free languages and correspondences}\label{CFcorrSec}

Recall the following definition of regular and context-free grammars.

\begin{defn}\label{RCFgr}
 Let $\fA$ be a finite alphabet, and let $\fA^*$ denote the set of all 
 strings of letters in $\fA$ of arbitrary (finite) length. It is a monoid under
 concatenation, with unit the empty word $\epsilon$, namely the {\em free monoid} generated by $\fA$.
 A {\em phrase structure grammar} is a set of the form $\cG=( \fQ, \fA, P, S)$ with
a finite set $\fQ$ of ``non-terminal symbols", the alphabet $\fA$ as the set
of  ``terminal symbols", a marked point $S$ (start symbol) in the set $\fQ$ 
and a finite set $P$ of {\em production rules}.
\begin{itemize}
\item The grammar is {\em regular} if all the production rules are of the form 
$A \to a$ or $A \to Aa$ (left-regular) for some $A\in \fQ$ and some $a\in \fA$
(or $A\to aA$ in the right-regular case). 
\item The grammar is {\em context-free} if the production
rules are of the form $A \to \alpha$ where $\alpha$ is a string of terminals and non-terminals.
\end{itemize}
The formal language $\cL=\cL_\cG$ computed by the grammar $\cG$ is the subset
$\cL \subset \fA^*$ of words in the terminal symbols 
obtained from a composition of production rules of $\cG$ starting with the non-terminal $S$. 
The syntactic monoid $M(\cL)\subset \fA^*$ of the formal language $\cL$ is the submonoid 
of $\fA^*$ generated by $\cL$. 
\end{defn}

In the theory of formal languages grammars are classified according to the Chomsky 
hierarchy \cite{Chomsky1}, \cite{Chomsky2} and a well known result about their 
machine recognition shows that formal languages are regular
if and only if they are computed by finite state automata, and context-free if 
and only if they are computed by pushdown stack automata. 

\smallskip

\begin{defn}\label{FSAPSA}
A finite state automaton (FSA) is a set  $\cM=(\fQ,F,\fA,\tau,q_0)$ with a finite set $\fQ$
of possible states, a subset $F\subset \fQ$ of final states and an initial state $q_0$, and
a set $\tau\subset \fQ\times \fA \times \fQ$ of transitions. A pushdown stack automaton (PSA) 
is a set $\cM=(\fQ,F, \fA, \Gamma, \tau, q_0, z_0)$ with $\fQ,F,\fA,q_0$ as in the FSA case
and with $\Gamma$ the stack alphabet and set of transitions 
$\tau\subset \fQ\times (\fA\cup \{ \epsilon \})\times \Gamma \times \fQ \times \Gamma^*$ and with an initial symbol $z_0\in \Gamma$ for the stack.
\end{defn}

In the case of an FSA one can interpret transitions $(q,a,q')$  as directed edges labelled 
by a letter $a\in \fA$ between two vertices $q, q'\in \fQ$ in a directed graph. This edge
describes going from state $q$ to state $q'$ when reading the letter $a\in \fA$ on an
instruction tape. In the case
of a PSA, a transition $(q,a,z,q',\gamma)$ describes the effect of reading the
letter $a\in \fA$ on an instruction tape and reading the letter $z\in \Gamma$ at the top
of the stack, moving from state $q$ to state $q'$, while depositing a string $\gamma\in \Gamma^*$
into the stack. The stack memory is accessible only last-in-first-out. 

\smallskip
\subsubsection{Rational correspondences}\label{RatCorrSec}

We give the following definition of {\em rational relation} 
between monoids (see \S 3 of \cite{Berstel} for other equivalent characterizations).

\begin{defn}\label{ratrel}
A {\em rational relation} (or rational correspondence) between two monoids $M,M'$ is
a subset $X\subset M\times M'$ such that there exists an alphabet $\fA$, 
two morphisms of monoids $\alpha: \fA^* \to M$ and $\beta: \fA^* \to M'$, and 
a regular language $\cL \subset \fA^*$ such that 
\begin{equation}\label{ratrelX}
X=\{ (\alpha(w), \beta(w))\in M\times M' \,|\, w\in \cL \} \, .  
\end{equation} 
We write $X=(\fA,\alpha,\beta,\cL)$ for such a relation. 
\end{defn}

A composition of rational relations can be defined as in \cite{Berstel} in the following way.

\begin{defn}\label{compXYdef}
given monoids $M,M',M''$ and rational correspondences $X\subset M\times M'$
and $Y\subset M'\times M''$, with $X=(\fA,\alpha,\beta,\cL)$ and $Y=(\fB, \alpha',\beta', \cL')$
as in Definition~\ref{ratrel}, where (possibly after a relabeling isomorphism) one can 
assume $\fA\cap \fB=\emptyset$. Let $Y\circ X \subset M\times M''$ be given by
\begin{equation}\label{corrcompose}
 Y\circ X=((\fA\cup \fB)^*, \alpha\circ \pi_\fA, \beta' \circ \pi_\fB, \hat\cL) \, , 
\end{equation} 
with $\hat\cL \subset \pi_\fA^{-1}(\cL) \cap \pi_\fB^{-1}(\cL')$ given by 
\begin{equation}\label{hatL}
\hat\cL=\{ w\in \pi_\fA^{-1}(\cL)\cap \pi_\fB^{-1}(\cL') \,|\, \beta\circ \pi_\fA(w) = \alpha'\circ\pi_\fB(w) \}\, ,
\end{equation}
with $\pi_\fA: (\fA\cup \fB)^*\to \fA^*$ and $\pi_\fB: (\fA\cup \fB)^*\to \fB^*$ the projection maps.
\end{defn}

It is known that this composition of rational relations is not always rational. A simple
example is given in Example~4.2 of \cite{Berstel}, with 
$M=\{ a \}^*$ and $M'=\{ b \}^* \times  \{ c \}^*$ 
and $M''=\{ b,c \}^*$ and with 
$$ X=(\fA=\{ a \}, \alpha={\rm id}, \beta(a)=(b,c), \cL=\fA^*)  $$
$$ Y=(\fB=\{ b, c \}, \alpha'(b)=(b,1), \alpha'(c)=(1,c), \beta'={\rm id}, \cL'=\fB^*) \, . $$
Then $Y\circ X$ is given by
$$ Y\circ X=( \fA\cup \fB, \alpha\circ\pi_\fA, \beta'\circ \pi_\fB, \hat\cL) $$
where the language $\hat\cL \subset \pi_\fA^{-1}(\cL)\cap \pi_\fB^{-1}(\cL')$ is given by
\eqref{hatL}.  
However, we see that the condition defining $\hat\cL$ gives 
$$ \{ w \,|\, \beta(\pi_\fA(w)) = \alpha'(\pi_\fB(w)) \}= \{ w\in \fB^*\,|\, \,  |w|_b = |w|_c \} $$
where $|w|_\ell$ is the number of occurrences of the alphabet letter $\ell$ in a word $w$. Thus, $\hat\cL\subset \fB^*$ is the set of
words with the same number of occurrences of the letters $b$ and $c$, and it is well known that this is not a regular language.
Thus, in this case $Y\circ X$ is {\em not rational}. 
This means that composition defined as in Definition~\ref{compXYdef} for rational correspondences
as in Definition~\ref{ratrel} is {\em not well defined}, as it does not preserve the rationality condition. However, we
can correct this problem, for the specific purpose of defining morphisms of context-free languages in the following way. 

\smallskip
\subsubsection{Rational transductions}\label{TransdSec}

We focus on a particular class of rational correspondences given by rational transductions, see \cite{Berstel}. 
The class of rational transductions we consider here is more restrictive than that of \cite{Berstel} since we
only consider transductions between free monoids $X\subset \fB^* \times \fC^*$. 

\begin{defn}\label{rattransd}
A {\em rational transduction} is a rational correspondences 
$$ X=(\fA,\alpha,\beta,\cL_{reg}) \, , $$
identified with a subset
$$ X=\{ (\alpha(w),\beta(w))\,|\, w\in \cL_{reg} \}  \subset \fB^* \times \fC^* \, ,  $$
for alphabets $\fB$ and $\fC$. 
A rational transdunction $X=(\fA,\alpha,\beta,\cL_{reg})\subset \fB^* \times \fC^*$,
maps subsets $\Omega \subset \alpha(\fA^*)\subset \fB^*$ to corresponding 
subsets $X(\Omega)\subset \fC^*$ by taking
\begin{equation}\label{XOmega}
X(\Omega)=\beta(\alpha^{-1}(\Omega)\cap \cL) \, . 
\end{equation}
\end{defn}

\smallskip

\begin{lem}\label{regXOmega}
Let $X=(\fA,\alpha,\beta,\cL_{reg})\subset \fB^* \times \fC^*$ be a rational
transduction. If $\Omega \subset \fB^*$ is a regular language, then $X(\Omega)\subset \fC^*$, defined as
in \eqref{XOmega} is also a regular language. If $\Omega$ is context-free then $X(\Omega)$ is also context free. 
\end{lem}

\proof Regular languages are stable under {\em homomorphisms of free monoids}, in the sense that if $\cR\subset \fA_1^*$ 
is a regular language and $\phi: \fA_1^* \to \fA_2^*$ and $\psi: \fA_0^* \to \fA_1^*$ are monoid homomorphisms, then
$\phi(\cR)\subset \fA_2^*$ and $\psi^{-1}(\cR)\subset \fA_0^*$ are regular languages, as can be seen in terms
of corresponding FSA. Thus, given $\Omega \subset \fB^*$ and $\alpha: \fA^* \to \fB^*$, we have
$\alpha^{-1}(\Omega)\subset \fA^*$ regular. The set of regular languages is also closed under intersection so
$\alpha^{-1}(\Omega)\cap \cL_{reg} \subset \fA^*$ is regular. This can also be seen in terms of FSA: if $\cR$ and
$\cR'$ are regular on the same alphabet $\fA$ and $\cM=(\fQ,F,\fA,\tau,q_0)$
and $\cM'=(\fQ',F',\fA,\tau',q_0')$ are FSAs computing $\cR$ and $\cR'$, respectively, then
$\cM''=(\fQ\times\fQ', F\times F', \fA, \tau'', (q_0,q_0'))$ with $((q_1,q_1'),a,(q_2,q_2'))\in \tau''$ if
$(q_1,a,q_2)\in \tau$ and $(q_1',a,q_2')\in \tau'$ is a FSA that computes
$\cR\cap \cR'$. Note then that if $\cR$ is a regular language on a given alphabet we can also view it as a regular
language on a larger alphabet (this can be seen in terms of FSAs), so that if $\cR$ and $\cR'$ have 
different alphabets we can see them as regular languages on $\fA\cup \fA'$.
The image $\beta(\alpha^{-1}(\Omega)\cap \cL_{reg} )\subset \fC^*$
under the homomorphism $\beta: \fA^* \to \fC^*$ is then also regular. Context-free languages are not
closed under intersection, but they are closed under intersection with a regular language, so
that intersecting a context-free language with $\cL_{reg}$ still gives a context-free language.
\endproof

\smallskip

As we mentioned above, regular and context-free languages are recognized (computed) by
FSA and PSA, respectively. In a similar way, rational transductions are computed by
automata, called {\em transducers} (see \S 6 of \cite{Berstel}). 

\begin{defn}\label{transducers}
A transducer is a set $\cT=(\fA,\fB, \fQ, F, \tau, q_0)$ consisting of an input alphabet $\fA$, an output alphabet $\fB$,
a set of states $\fQ$ with an initial state $q_0$ and a set $F$ of final states and a set of transitions
$\tau \subset \fQ \times \fA^* \times \fB^* \times \fQ$. 
\end{defn}

Let $\cP(q,q')$ denote the set of paths $(q_1,a_1,b_1,q_2)\cdots (q_n,a_n,b_n,q_{n+1})$ with $q=q_1$,
$q'=q_{n+1}$ and $(q_i,a_i,b_i,q_{i+1})\in \tau$, and let $\cP(q_0,F)=\cup_{q\in F} \cP(q_0,q)$. 
The sets $\cP(q,q')$ are regular languages and so is $\cP(q_0,F)$ and every rational transduction can be
obtained in this way (see Theorem~6.1 of \cite{Berstel}).

\smallskip
\subsection{Category of context-free languages}

We now show that we can use rational transductions to construct a category of context-free languages.

\begin{thm}\label{catFL}
Context-free formal languages form a category  $\cC\cF\cL$ where the objects are context-free languages $\cL$
and where the non-identity morphisms $X \in Mor_{\cC\cF\cL}(\cL, \cL')$ of two context-free languages $\cL\subset \fB^*$
and $\cL'\subset \fC^*$ are rational transductions 
$X=(\fA,\alpha,\beta,\cL_{reg})\subset \fB^*\times \fC^*$, with $\cL \subset \alpha(\fA^*)$
and $\cL'\subset \beta(\fA^*)$, and such that $X(\cL)\subset\cL'$.
\end{thm}

\proof We need to show that, for any $\cL,\cL', \cL''$ context free, the composition of morphisms 
$\circ: Mor_{\cC\cF\cL}(\cL, \cL') \times Mor_{\cC\cF\cL}(\cL', \cL'')$ is well defined and associative.
Up to a possible relabeling isomorphism, we can assume for simplicity that the alphabets $\fB, \fC, \fD$ 
of the three languages are disjoint. 
While we have seen that composition of rational correspondences of monoids is not in general well
defined, composition of rational transduction is well defined (see Theorem~4.4 of \cite{Berstel}). 
In our case, we consider as in Definition~\ref{compXYdef} 
the composition $Y\circ X$ of $X=(\fA,\alpha,\beta,\cL_{reg}) \subset \fB^* \times \fC^*$ and 
$Y=(\fA',\alpha',\beta',\cL'_{reg})\subset \fC^*\times \fD^*$. First observe that 
$$ \beta(\cL_{reg})\cap \alpha'(\cL'_{reg}) \subset \fC^*  $$
is a regular language by closure of the set of regular languages under homomorphisms 
{\em of free monoids} and under intersections. 
Closure under homomorphisms of free monoids and intersections also give that 
$\pi_{\fA}^{-1}(\cL_{reg})\cap \pi_{\fA'}^{-1}(\cL'_{reg}) \subset (\fA\cup \fA')^*$ is a regular
language and the subset $\hat\cL$ defined as in \eqref{hatL} is given by 
\begin{equation}\label{hatL2}
 \hat\cL =(\beta\pi_{\fA})^{-1}( \beta(\cL_{reg})\cap \alpha'(\cL'_{reg}) ) = (\alpha'\pi_{\fA'})^{-1} ( \beta(\cL_{reg})\cap \alpha'(\cL'_{reg}) ) 
 \subset (\fA\cup \fA')^* \, , 
\end{equation} 
and is also a regular language, again by stability under homomorphisms of free monoids. We
identity $Y\circ X$ with the set
\begin{equation}\label{corrcompose2}
Y\circ X=\{ (\alpha\circ\pi_\fA(w) ,\beta'\circ\pi_{\fA'}(w))\,|\, w\in \hat\cL \} \subset \fB^* \times \fD^*  
\end{equation} 
Moreover, if $X(\cL)=\beta(\alpha^{-1}(\cL)\cap \cL_{reg})\subset \cL'$
and $Y(\cL')=\beta'({\alpha'}^{-1}(\cL')\cap \cL'_{reg})\subset \cL''$, then 
$$ Y(X(\cL)) = \beta'({\alpha'}^{-1}(\beta(\alpha^{-1}(\cL)\cap \cL_{reg})) \cap \cL'_{reg}) \subset \cL''  \, . $$
We can identify this with
$$  Y(X(\cL)) = (\beta' \circ \pi_{\fA'}) ( (\alpha\circ \pi_{\fA})^{-1} ( \cL) \cap \hat\cL )  \, ,  $$
since both sets are equal to the pairs $(w,w')\in (\pi_{\fA}, \pi_{\fA'})((\fA\cup \fA')^*)=\fA^*\times {\fA'}^*$
with $w\in \cL_{reg}$ and $w'\in \cL'_{reg}$ and with $\beta(w)=\alpha'(w')$ and $\alpha(w)\in \cL$. 
This shows that the composition of morphisms is well defined. 
For associativity, consider rational transductions
$X_i=(\fA_i,\alpha_i,\beta_i,\cL_i)\subset \fB_i^*\times \fB_{i+1}^*$ for $i=1,2,3$, where again after a possible
relabeling isomorphism we can assume $\fA_i \cap \fA_j=\emptyset$ and $\fB_i\cap \fB_j=\emptyset$ 
for $i\neq j$. To see  that we have $(X_1 \circ X_2)\circ X_3 = X_1\circ (X_2\circ X_3)$ we use the notation
$\hat\cL_{i,i+1} \subset \pi_{\fA_i}^{-1}(\cL_i) \cap \pi_{\fA_{i+1}}^{-1}(\cL_{i+1})$ and 
$\hat\cL_{(12)3}\subset \pi_{\fA_1\cup \fA_2}^{-1}(\hat\cL_{12})\cap \pi_{\fA_3}^{-1}(\cL_3)$ and
$\hat\cL_{1(23)}\subset \pi_{\fA_1}^{-1}(\cL_1)\cap \pi_{\fA_2\cup \fA_3}^{-1}(\hat\cL_{23})$ and 
observe that we have
$$ \hat\cL_{(12)3}=\left\{ w\,|\, \begin{array}{l} \alpha_2\circ\pi_{\fA_2}\circ \pi_{\fA_2\cup \fA_3}(w)=\beta_1\circ \pi_{\fA_1}(w) \\
\beta_2\circ \pi_{\fA_2}(w)=\alpha_3\circ \pi_{\fA_3}(w) \end{array}
\right\} $$
$$ =\left\{ w\,|\, \begin{array}{l} 
\alpha_2\circ\pi_{\fA_2}(w)=\beta_1\circ \pi_{\fA_1}\circ \pi_{\fA_1\cup \fA_2}(w)  \\
\beta_2\circ \pi_{\fA_2} \circ \pi_{\fA_1\cup \fA_2} (w) = \alpha_3 \circ \pi_{\fA_3}(w) \end{array}
  \right\} = \hat\cL_{1(23)} \, . $$
\endproof

\smallskip
\subsubsection{D0L-systems and inflation rules}\label{D0Lsec}

A particular type of formal languages has proved especially useful to model the geometry
of quasicrystals, the D0L-systems considered for this purpose in \cite{GarEsc1}. These
are a special case of the general $L$-systems (Lindenmayer systems, see \cite{RozSal} for a detailed presentation).
The main aspect of $L$-systems that distinguishes them from the generative grammars
of formal languages in the sense we recalled above is that the production rules in this
case apply ``in parallel" wherever they can be applied, rather than sequentially. The
subclass of D0L-systems have the property that the generative process starts with
a single initial word and continues by applying context-free rules at each step. 
More precisely, D0L-systems and D0L-languages are defined as follows.

\begin{defn}\label{D0Lsys}
A D0L-system is a triple $\cD=(\Sigma, \Phi, \Omega)$ where $\Sigma$ is a finite alphabet,
a monoid homomorphism $\Phi: \Sigma^* \to \Sigma^*$ and an initial word $\Omega \in \Sigma^*$.
The D0L-language generated by this system is
\begin{equation}\label{D0Llang}
\cL_\cD=\{ \Phi^n(\Omega) \,|\, n\in \N \} \, . 
\end{equation}
\end{defn}

We show here that it is more natural to consider D0L-systems and D0L-languages not as
a special type of formal languages but as a special case of correspondences of 
formal languages. This leads naturally to a generalization of the setting of D0L-systems 
(and some more general classes of $L$-systems) and their applications to modeling
aperiodic geometries and the associated physical systems. 

\smallskip

\begin{prop}\label{D0lcorr}
Given a homomorphism of free monoids $\Phi: \fB^* \to \fC^*$ and a pair
of a regular language $\cL \subset \fB^*$ and and a context-free language $\cL'\subset \fC^*$, such
that $\Phi(\cL)\subset \cL'$ determines a morphism $X_\Phi\in  {\rm Mor}_{\cC\cF\cL}(\cL,\cL')$
in the category $\cF\cL$ of formal languages of Theorem~\ref{catFL}.
In particular, a D0L-system $\cD=(\Sigma, \Phi, \Omega)$ determines a semigroup 
\begin{equation}\label{semigrPhi}
\cX_{\Phi} \subset {\rm End}_{\cC\cF\cL}(\Sigma^*) \, ,
\end{equation}
of endomorphisms in the category $\cC\cF\cL$, with
$$ \cX_\Phi = \cup_n X_{\Phi^n} \, . $$
For $\cL=\{ \Omega \}$ the regular language given by a singleton, the
image under the semigroup \eqref{semigrPhi} is the D0L-language
\begin{equation}\label{D0Lcorr}
 \cX_{\Phi}(\Omega) =\cL_\cD \, . 
\end{equation}
\end{prop}

\proof Given a monoid homomorphism $\Phi: \fB^* \to \fC^*$ with $\Phi(\cL)\subset \cL'$, 
consider the graph
$$ X_\Phi =\{ (w,\Phi(w)) \,|\, w\in \cL \} \subset \fB^*  \times \fC^* \, . $$ 
If $\cL$ is a regular language, we can take $\alpha={\rm id}$ and $\beta=\Phi$ in
\eqref{ratrelX} and we obtain a rational transduction. Clearly when $\cL'=\cL$ all the
iterations $\Phi^n$ are also correspondences given by rational transductions and the composition
as monoid homomorphisms matches the composition as correspondences, since $\hat\cL$ as
in \eqref{hatL2} is just $\Phi\circ\Phi(\cL)$ in this case. 
For a D0L-system $\cD=(\Sigma, \Phi, \Omega)$, the freely generated
monoid $\Sigma^*$ on the alphabet $\Sigma$ is computed by a FSA consisting of
a single vertex and an edge for each letter in $\fA$. Thus, the monoid
endomorphisms $\Phi^n: \Sigma^* \to \Sigma^*$ define rational transduction correspondences
$X_{\Phi^n} \in {\rm Mor}_{\cC\cF\cL}(\Sigma^*,\Sigma^*)$ so that 
$\cX_\Phi = \cup_n X_{\Phi^n}$ is a semigroup under composition.
We then obtain \eqref{D0Lcorr} from \eqref{D0Llang}.
\endproof

\section{Formal languages and spin systems}\label{FLspinSec}

We now consider aperiodic spin chains as obtained from formal languages
via a functorial construction based on the category constructed in Theorem~\ref{catFL}.
This formalism will incorporate the well known cases of substitution rules and of
block spin renormalization.

\subsection{Spin chains}

Constructing aperiodic spin chains and aperiodic 2D Ising models using substitution 
rules encoded by formal languages, specifically D0L-systems, is a well established technique, see \cite{GriBaa}. 
We give here a brief review some aspects of aperiodic spin chains and we
present a setting that will incorporate these into a construction compatible with
the category of context-free formal languages of Theorem~\ref{catFL}.

\smallskip 

A one-dimensional Ising chain has spin variables $\sigma_j =\pm 1$ and a Hamiltonian
\begin{equation}\label{1Dspinch}
H(\sigma) = -\sum_{j=1}^N J_j \sigma_j \sigma_{j+1} - \sum_{j=1}^N H_j \, \sigma_j \, ,
\end{equation}
where the sum takes place over a (finite but large) set of sites, seen as points on the
real line (one can assume periodic boundary conditions $\sigma_N=\sigma_1$). 
The Boltzmann weight of a spin configuration is defined as
\begin{equation}\label{BoltzW}
W(\sigma)=\prod_{j=1}^N W_j(\sigma_j, \sigma_{j+1})=e^{-\beta \, H(\sigma)}   =\prod_{j=1}^N e^{\beta (J_j  \sigma_j \sigma_{j+1}  +\frac{1}{2}(H_j \sigma_j + H_{j+1} \sigma_{j+1}))}  \, ,
\end{equation}
with $\beta$ the inverse temperature (up to the Boltzmann constant for dimensionality), so
that the partition function is given by
\begin{equation}\label{ZNpartition}
Z_N(\beta)=\sum_\sigma e^{-\beta H(\sigma)} \, . 
\end{equation}
One can assume for simplicity the case of constant magnetization $H_j=H$, though for now we will keep $H_j$ variable.

\smallskip

If we associate to each site $i$ of the spin chain a $2\times 2$ {\em transfer matrix}
$$ T_j =\begin{pmatrix} e^{\beta (J_j + \frac{1}{2}(H_j+H_{j+1}))} & e^{-\beta ( J_j - \frac{1}{2}(H_j-H_{j+1}))} \\
e^{-\beta (J_j + \frac{1}{2}(H_j-H_{j+1}))} & e^{\beta (J_j - \frac{1}{2}(H_j+H_{j+1}))}
\end{pmatrix} $$
then we can equivalently write 
$$ Z_N(\beta)= \Tr ( T_1 \cdots T_N ) =\Tr(T) \, , $$
where $T=T_1 \cdots T_N$ is the {\em monodromy matrix} of the spin chain (see \S 2.1 of \cite{GriBaa}). 

\smallskip

A well known technique in spin chain models is Kadanoff's renormalization. 
Consider a site $i\in \{ 1,\ldots, N \}$ of a spin chain of length $N$ sites 
(and periodic conditions). Consider a transformation that integrates out 
the degrees of freedom at the $i$-th site (namely we add over the values 
$\sigma_i=\pm 1$) while maintaining the same partition function of the system
with a new Hamiltonian for the remaining degrees of freedom that is of the
same form up to a change in the coupling constants. Assume that the factor in the
partition function that involves the $i$-th site is given by
\begin{equation}\label{Zifactor}
 \exp( K  (\sigma_{i-1} \sigma_i + \sigma_i \sigma_{i+1}) + h \sigma_i + \frac{h}{2}(\sigma_{i-1} + \sigma_{i+1})) ) \, , 
\end{equation} 
with $K=\beta J$ and $h=\beta H$.
Summing over the values $\sigma_i=\pm 1$ gives
$$ \exp( (K+\frac{h}{2}) (\sigma_{i-1} + \sigma_{i+1}) +h  ) + \exp((-K+\frac{h}{2})(\sigma_{i-1} + \sigma_{i+1}) - h  ) \, . $$
We require that this term should be of the form
$$ \exp( K' \sigma_{i-1}\sigma_{i+1} +\frac{h'}{2} ( \sigma_{i-1} + \sigma_{i+1})  + 2g )  $$
up to an overall factor $\exp(2 g)$ that has no effect on the dynamics as it is independent of the spins
$\sigma$ (we can always assume the Hamiltonian is defined up to a global translation).
This leads to the well known renormalization transformation
\begin{equation}\label{rentransf}
\begin{array}{ll}
 K' = & \frac{1}{4} \log \frac{\cosh(2K+h) \cosh(2K-h))}{\cosh^2(h)} \\[3mm]
 h' = & h +\frac{1}{2} \log\frac{\cosh(2K+h)}{\cosh(2K-h)} \\[3mm]
 g = &\frac{1}{4} \log(\cosh(2K+h) \cosh(2K-h) \cosh(h)) 
\end{array}
\end{equation}
This gives rise to a transformation $(K,h)\mapsto (K',h')$ that maps
the original Hamiltonian $H(\sigma)$ to the new Hamiltonian $H'(\hat\sigma)$
where $\hat\sigma=(\sigma_1,\ldots,\hat \sigma_i,\ldots,\sigma_N)$ with the
spin $\sigma_i$ removed, and the above change of parameters, and  with
$$ Z_N(\beta)=\sum_\sigma e^{-\beta H(\sigma)} = \sum_{\hat\sigma} e^{-\beta H'(\hat\sigma)} 
=Z'_{N-1}(\beta) \, , $$
up to an overall multiplicative factor $\exp(2g)$ as above.

\smallskip

The transformation above is obtained under the simplifying assumptions that
$J_i=J_{i+1}=J$ and that $H_i=H$. In fact for simplicity, we will further assume
here that the external magnetization $H \equiv 0$ so that the above transformation
simplifies to
\begin{equation}\label{rentransfH0}
\begin{array}{ll}
 K' = & \frac{1}{2} \log \cosh(2K)  \\[3mm]
 g = &\frac{1}{2} \log(2 \cosh(2K)) \, .
\end{array}
\end{equation}

We consider the more general case where
a similar procedure of summing out one of the degrees of freedom $\sigma_i$ of
the spin chain is performed with $J_i=J_a \neq J_{i+1}=J_b$ (while we still assume
that the external magnetization is trivial $H_i=H=0$). 

\begin{lem}\label{renJab}
For $J_i=J_a \neq J_{i+1}=J_b$ the renormalization transformation (spin decimation) that sums out
the $\sigma_i$ degrees of freedom while maintaining the same Hamiltonian up to
a constant additive shift give the new parameters 
\begin{equation}\label{rentransfH0Jab}
\begin{array}{ll}
 K' = & \frac{1}{2} \log\left( \frac{ \cosh(K_a+K_b) }{\cosh(K_a-K_b)} \right)  \\[3mm]
 g = &  \frac{1}{4} \log(\cosh(K_a+K_b) \cosh(K_a-K_b) )  \, .
\end{array}
\end{equation}
\end{lem}

\proof
Summing over the values $\sigma_i=\pm 1$, with the Hamiltonian 
$$  \exp( K_a  \sigma_{i-1} \sigma_i + K_b \sigma_i \sigma_{i+1}  
+ h \sigma_i + \frac{h}{2}(\sigma_{i-1} + \sigma_{i+1})) ) $$
(generalizing \eqref{Zifactor} to include the case $K_a\neq K_b$) we obtain 
$$ \exp( (K_a +\frac{h}{2}) \sigma_{i-1} + (K_b +\frac{h}{2})  \sigma_{i+1}) + h  ) 
+ \exp((-K_a+\frac{h}{2})\sigma_{i-1} +(-K_b+\frac{h}{2}) \sigma_{i+1}) - h  ) $$
We then have, for the possible values $\pm 1$ of $\sigma_{i\pm 1}$ the equations
$$ \exp( K_a + K_b + 2h  ) + \exp( - (K_a +K_b))  =\exp( K' + h' +2g ) $$
$$ \exp( K_a  - K_b  + h  )  + \exp(-K_a + K_b - h  ) = \exp (- K' + 2g) $$
$$ \exp( -K_a   + K_b  + h  )  + \exp( K_a -K_b - h  ) =\exp( - K' +2g ) $$
$$ \exp( -(K_a + K_b) ) + \exp( K_a + K_b - 2h  )  =\exp (K' - h' +2g)\, .  $$
Under the assumption that $h=0$ these reduce to
$$ \exp(K_a+K_b)+\exp(-(K_a+K_b)) =\exp(K') e^{2g} $$
$$ \exp(K_a-K_b)+\exp(-(K_a-K_b)) = \exp(-K') e^{2g} \, , $$
which are solved by \eqref{rentransfH0Jab}.
\endproof

\smallskip

Since the factor $\exp(2g)$ has no effect on the dynamics, we just consider this as a
transformation
\begin{equation}\label{RenTransf}
J=\{ J_j \}_{j \in \cI} \mapsto \cR(J)=\{ J'_j \}_{j\in \cI'}
\end{equation}
where $\beta J'=K'$ as in \eqref{rentransfH0Jab} with $\cI'=\cI \smallsetminus \{ i \}$.
We still write $J\mapsto \cR(J)=J'$ for the iterated application of the transformation
\eqref{rentransfH0Jab} that decimates the spins in a subset $\cJ\subset \cI$, with
$\cI'=\cI\smallsetminus \cJ$. In particular, we write $J'_{a'} =\cR(J_w)$ with $w=ab$
if $J_{a'}$ is the new single interaction term that replaces the previous pair $J_a, J_b$
under \eqref{rentransfH0Jab}. 

\smallskip

We also make the following simple observation that will be useful in the following.

\begin{rem}\label{decouple} {\rm
If we allow some of the $J_i=0$, under the assumption that also $H=0$, then
the spin chain decouples into two non-interacting spin chains, one ending at
$\sigma_i$ and one starting at $\sigma_{i+1}$. The partition function correspondingly
factors as a product of the partition functions of these independent subsystems. }
\end{rem} 

\subsection{A category of spin chains} \label{spincatSec}

We organize spin chains with renormalization transformations into
a categorical structure.

\begin{defn}\label{SpinSetDef}
Consider sets of the form $\Upsilon=(\fA, \Omega, J_\fA)$, where
$\fA$ is a finite alphabet, $\Omega\subset \fA^*$ is a nonempty set of words
and $J_{\fA}=\{ J_a \}_{a\in \fA}$ is a set of parameters with $J_a\geq 0$
for all $a\in \fA$. We define a {\rm spin chain set} $S_\Upsilon$ by taking
\begin{equation}\label{SUps}
 S_\Upsilon :=\sqcup_{w\in \Omega} S_w\, , 
\end{equation} 
where $S_w$ is a spin chain of length $\ell(w)$, the length of the word $w$,
with Hamiltonian \eqref{1Dspinch} with $J_j=J_a$ if the word
$w$ has the $j$-th letter $w_j=a\in \fA$, and with all $H_j=H=0$. 
\end{defn}

We need the following observation for the next definition.

\begin{lem}\label{homdecomp}
Let $\alpha: \fA^* \to \fB^*$ be a homomorphism of free monoids generated by 
the respective finite alphabets $\fA$ and $\fB$. Then we can uniquely decompose
$\fA=\fA_\alpha \sqcup \fA_{\alpha^\perp}$ where $\alpha(a)\neq \epsilon$ (the
unit of the monoid $\fB^*$) for all $a\in \fA_\alpha$ and $\alpha(a) = \epsilon$
for all $a\in  \fA_{\alpha^\perp}$.  
\end{lem}

\proof the homomorphism $\alpha: \fA^* \to \fB^*$ is completely determined by
the words $\alpha(a)\in \fB^*$ for the letters $a\in \fA$. The kernel ${\rm Ker}(\alpha)=\{ w\in \fA^*\,|\, \alpha(w)=\epsilon \}$ consists of all the words $w=a_1\ldots a_n$ where each letter $a_i$
has $\alpha(a_i)=\epsilon$, since $\fB^*$ is the free monoid on the set $\fB$, so a word
$\alpha(w)=\epsilon \in \fB^*$ cannot have any subword $\alpha(w)=w' w''$ with either
$w' \neq \epsilon$ or $w''\neq \epsilon$, as that would be a nontrivial relation. Thus, since
$\alpha(w)=\alpha(a_1)\ldots \alpha(a_n)$, each $\alpha(a_i)=\epsilon$. This 
${\rm Ker}(\alpha)$ with a free submonoid of $\fA^*$ generated by the subset 
$\fA_{\alpha^\perp}\subset \fA$ of letters $a\in \fA$ with $\alpha(a)=\epsilon$. 
\endproof

\begin{defn}\label{splitchain}
Given $\alpha: \fA^* \to \fB^*$ a homomorphism of free monoids, and let $\Pi$ be a choice of a decomposition 
$\fA=\fA_1 \sqcup \cdots \sqcup \fA_\ell$ of $\fA$, with $\Pi_\alpha$ the decomposition 
$$ \fA =\bigsqcup_i \fA_{i \alpha} \sqcup \bigsqcup_i \fA_{i \alpha^\perp}$$ 
given by the intersection of the decomposition $\Pi$ with the decomposition
$\fA=\fA_\alpha \sqcup \fA_{\alpha^\perp}$
of Lemma~\ref{homdecomp}.  Consider an alphabet $\fA_o=\fA \cup \{ o \}$
and extend the homomorphism to $\alpha_o: \fA_o^* \to \fB^*$ by setting $\alpha_o(o)=\epsilon$
and $\alpha_o(a)=\alpha(a)$ for all $a\in \fA$. 
Given a set $\Omega \subset \alpha(\fA^*)\subset \fB^*$, a {\rm disconnection of $\Omega$ at $\Pi_\alpha$},
denoted by $\alpha^{-1}(\Omega)_o$, is a subset of $\alpha_o^{-1}(\Omega)$
obtained by inserting the letter $o$ into the words of $\alpha^{-1}(\Omega)$ by
the rule $uv \mapsto uov$, whenever the adjacent letters $u,v$ 
are in different pieces of the decomposition $\Pi_\alpha$ of $\fA$.
The letter $o$ used for disconnection can be a letter already present in $\fA_{\alpha^\perp}$ or a new letter. 
\end{defn}

 We can then give the following definition.

\begin{defn}\label{commdecima}
Given two spin chain sets $S_\Upsilon$ and $S_{\Upsilon'}$ as in
Definition~\ref{SpinSetDef}, with $\Upsilon=(\fB, \Omega, J'_\fB)$
and $\Upsilon'=(\fC, \Omega', J_\fC)$.
A common {\em decimation and disconnection} of both is a spin chain set $S_{\hat\Upsilon}$ 
with $\hat\Upsilon=(\fA_o, \hat\Omega, \hat J_{\fA_o})$, such that there are homomorphisms
of free monoids $\alpha: \fA^* \to \fB^*$ and $\beta: \fA^* \to \fC^*$ such
that $\Omega\subset \alpha(\fA^*)$ and $\Omega'\subset \beta(\fA^*)$, with
$\hat \Omega \subset \alpha^{-1}(\Omega)_o$ and $\beta_o(\hat\Omega)\subset \Omega'$,
where $\alpha^{-1}(\Omega)_o$ is the disconnection of $\Omega$ obtained as in Definition~\ref{splitchain} 
with respect to the decomposition
$$ \fA=\fA_{\alpha\beta}\sqcup \fA_{\alpha\beta^\perp}\sqcup \fA_{\alpha^\perp \beta}\sqcup \fA_{\alpha^\perp \beta^\perp} $$
where the pieces of this decomposition come from the intersections of the pieces of the two decompositions
$\fA=\fA_\alpha \sqcup \fA_{\alpha^\perp}$ and $\fA=\fA_\beta\sqcup \fA_{\beta^\perp}$, as in Lemma~\ref{homdecomp}. Then $\hat\Upsilon$ has
\begin{equation}\label{hatJa}
 \hat J_a =\left\{ \begin{array}{ll} \cR(J_{\alpha(a)}) = \cR(J_{\beta(a)}) & a\in \fA_{\alpha\beta} \\
 \cR(J_{\alpha(a)}) & a\in \fA_{\alpha\beta^\perp} \\
 \cR(J_{\beta(a)}) & a\in  \fA_{\alpha^\perp \beta} \\
 0 & a\in \fA_{\alpha^\perp \beta^\perp}  \, . 
\end{array}\right. 
\end{equation}
We also take $ \hat J_o=0$.
The equality $\cR(J_{\alpha(a)}) = \cR(J_{\beta(a)})$ for $a \in \fA_{\alpha\beta}$ is a
condition on the existence of a common decimation. We write the relation \eqref{hatJa}
as 
$$ \hat J_{\fA}=\cR_\alpha (J_{\fB}) =\cR_\beta(J_{\fC}) \, . $$ 
We then write 
$$ \bD(S_{\Upsilon'},S_\Upsilon)=(S_{\hat\Upsilon},\alpha,\beta) $$
for such a common decimation of $S_\Upsilon$ and $S_{\Upsilon'}$.
\end{defn}
\smallskip

\begin{rem}\label{clarifydecima} {\rm
Note that the operation descrined in Definition~\ref{commdecima}
 involves both a  ``common decimation" or ``common renormalization" 
and a combination of disconnected uncoupled systems.
The part of the construction coming
from $\fA_{\alpha\beta}$ contributes a common decimation, while the part from
$\fA_{\alpha\beta^\perp}\sqcup \fA_{\alpha^\perp \beta}$
contributes a combination of spin chains as independent decoupled 
systems (see Remark~\ref{decouple} above), with the decoupling realized
by the disconnection $\alpha^{-1}(\Omega)_o$. In particular, in cases where
$\fA_{\alpha\beta}=\emptyset$, this operation is just a combination of
independent systems. Including both decimation and combination of
independent subsystems is necessary, in order to obtain good compositional properties
for these correspondences, as we discuss in Proposition~\ref{SpinCat} below.
}\end{rem}

\smallskip

\begin{prop}\label{SpinCat}
Let ${\rm Obj}(\cS)$ be the set of spin chain sets $S_\Upsilon$
as in \eqref{SUps} with $\Upsilon$ as in Definition~\ref{SpinSetDef} and
with ${\rm Mor}_\cS(S_\Upsilon, S_{\Upsilon'})$ given by common
decimations and disconnections of $S_\Upsilon$ and $S_{\Upsilon'}$, as in Definition~\ref{commdecima}.
The sets  ${\rm Obj}(\cS)$ and ${\rm Mor}_\cS(S_\Upsilon, S_{\Upsilon'})$ 
define a category $\cS$. 
\end{prop}

\proof Every object $S_\Upsilon$ has a unique identity morphism consisting
of the trivial decimation that keeps all the sites intact, namely $S_\Upsilon$ itself.
We need to construct the composition of morphisms 
$$ \bD(S_{\Upsilon''},S_{\Upsilon'}) \circ \bD(S_{\Upsilon'},S_\Upsilon) $$
which we also write as $S_{\hat\Upsilon'}\circ S_{\hat \Upsilon}$, with
$\bD(S_{\Upsilon'},S_\Upsilon')=(S_{\hat\Upsilon},\alpha,\beta)$ a common 
decimation and disconnection of $S_\Upsilon$ and $S_{\Upsilon'}$ and 
$\bD(S_{\Upsilon''},S_{\Upsilon'}) =(S_{\hat\Upsilon'},\alpha',\beta')$ 
a common decimation and disconnection of $S_{\Upsilon'}$ and $S_{\Upsilon''}$. 
The composition $S_{\hat\Upsilon'}\circ S_{\hat \Upsilon}$ is of the form
$$ \bD(S_{\Upsilon''},S_\Upsilon)= (S_{\hat\Upsilon''}, \alpha\circ \pi_{\fA}, \beta'\circ \pi_{\fA'}) \, , $$ 
with $\hat\Upsilon''=((\fA\sqcup \fA')_o, \hat\Omega'', \hat J''_{(\fA\sqcup \fA')_o})$,
where $\hat\Upsilon=(\fA_o,\hat\Omega,\hat J_{\fA_o})$ and $\hat\Upsilon'=(\fA_o', \hat\Omega',\hat J'_{\fA'_o})$
as in  Definition~\ref{commdecima}, and $\hat\Omega''$ and $\hat J''_{\fA\sqcup \fA'}$ obtained
in the following way:
\begin{equation}\label{composedecima}
\hat\Omega''=\{ w\in \pi_{\fA}^{-1}(\hat\Omega)_o \cap \pi_{\fA'}^{-1}(\hat\Omega')_o \,|\, \beta_o(\pi_\fA(w))=\alpha_o'(\pi_{\fA'}(w)) \} \, , 
\end{equation}
\begin{equation}\label{composeJdecima}
\hat J''_a =\hat J_a\, \, \forall a\in \fA \ \ \text{ and } \ \  \hat J''_{a'} =\hat J'_{a'}\, \, \forall a'\in \fA' \, ,
\end{equation}
with $\hat J''_o=0$. 
The argument for associativity of the composition defined in this way follows closely the
analogous argument in Theorem~\ref{catFL}, so we omit the details here.
\endproof

\subsection{Context-free languages and aperiodic spin chains}\label{CFLspinSec}

Suppose given a regular or more generally context-free grammar $\cG$.
To any word $w\in \cL_\cG$ in the associated language, of length $\ell(w)=N$
in the alphabet $\fA$, we can assign a spin chain as in \eqref{1Dspinch} with
coupling constants $J_a$ for each letter $a\in \fA$, where $J_j=J_a$ if the word
$w\in \cL_\cG$ has $j$-th letter $w_j=a\in \fA$. This assignment of spin chain
sets to context-free languages is functorial with respect to the categories
described in Theorem~\ref{catFL} and Proposition~\ref{SpinCat}.

\begin{thm}\label{CFspinFunctor}
Let $\cC\cF\cL\cJ$ be the category whose objects are pairs $(\cL,J)$ with
$\cL \subset \fB^*$ a context-free language (an object in $\cC\cF\cL$)
and $J: \fB \to \R_+^{\# \fB}$ a function, where we write $J_\fB=J(\fB)$. Nontrivial morphisms 
$(X,\hat J)\in {\rm Mor}_{\cC\cF\cL\cJ}((\cL,J),(\cL',J'))$, with $\cL\subset \fB^*$ and $\cL'\subset \fC^*$, 
are pairs of a morphism $X=(\fA,\alpha,\beta,\cL_{reg}) \in  {\rm Mor}_{\cC\cF\cL}(\cL,\cL')$ 
given by a rational transduction, and
$\hat J: \fA \to \R_+^{\# \fA}$ defined as in \eqref{hatJa}. In particular, such nontrivial morphisms
in ${\rm Mor}_{\cC\cF\cL\cJ}((\cL,J),(\cL',J'))$ exist under the compatibility condition of \eqref{hatJa}
between $J_\fB$ and $J'_\fC$. 
There is a functor $\fS: \cC\cF\cL\cJ \to \cS$ with assigns to a pair $(\cL,J)$ of a
context-free language $\cL\subset \fA^*$ on an alphabet $\fA$, and a set $J_\fA$ 
the spin chain set
$\fS(\cL)=S_\Upsilon$ with $\Upsilon=(\fA, \cL, J_\fA)$ and to a 
morphism $(X,\hat J) \in {\rm Mor}_{\cC\cF\cL\cJ}((\cL,J),(\cL', J'))$ 
it assigns a common decimation and disconnection
$\bD(S_{\Upsilon'},S_\Upsilon)=(S_{\hat\Upsilon},\alpha,\beta)$
where for $X=(\fA, \alpha,\beta, \cL_{reg})$, identified with the set
$X=(\alpha\times\beta)(\cL_{reg}) \subset \fB^*\times \fC^*$, 
the chain set $S_{\hat\Upsilon}$ has
$$ \hat\Upsilon=(\fA, X\cap (\cL\times \cL'), \hat J_{\fA})\, , $$
with $\hat J_{\fA}=\cR_\alpha (J_{\fB})=\cR_\beta(J_{\fC})$.
\end{thm}

\proof The composition of morphisms in $\cC\cF\cL\cJ$ is well defined and
associative, by the same argument as in Proposition~\ref{SpinCat} and 
Theorem~\ref{catFL}. 
The compatibility of $\fS$ with the composition of morphisms follows from the
compatible definitions of composition in \eqref{composedecima} and
\eqref{composeJdecima} and in Theorem~\ref{catFL}.
\endproof

\subsection{Multiple context free grammars} \label{MCFGsec}

A well known generalization of context-free grammars to a class of mildly-context-sensitive grammars
is given by the {\em multiple context-free grammars} (MCFG), introduced in \cite{Kasa}. Unlike
ordinary context-free grammars, MCFGs manipulate tuples of strings rather than individual strings
of terminals and nonterminals. We recall the definition of MCFGs as follows. Slightly
different but equivalent definitions are in use in the literature.

\begin{defn}\label{MCFGdef}
An {\em m-multiple context-free grammar} (m-MCFG) is a set $\cG=(\fA, \cN, P, S)$
consisting of an alphabet $\fA$ of terminal symbols, a set $\cN$ of nonterminals,
where each $N\in \cN$ has an associated dimension $d(N)$ with $1\leq d(N)\leq m$,
an initial nonterminal $S\in \cN$ of dimension $d(S)=1$, and a set $P$ of production
rules of the form
\begin{equation}\label{prodrulesMCFG}
A \mapsto f(B_1, \ldots, B_q) \ \ \ \  \text{ or } \ \ \ \  A \mapsto (w_1, \ldots, w_{d(A)}) \, ,
\end{equation} 
where the $w_k\in \fA^*$ are strings of terminals, and 
where $f: \cN^q \to \cN^{d(A)}$ are functions with $f=(f_1,\ldots, f_{d(A)})$ and each
$\omega_k=f_k(B_1, \ldots, B_q)$ 
is a string involving terminal symbols and the components $x_{j,r}$ of a subset of the nonterminals 
$B_j=(x_{j,r})_{r=1}^{d(B_j)}$, with each $B_j$ occurring at most once in the string 
$\omega_k$. 
The language $\cL_\cG$ consists of all the strings of terminals that are obtained starting from 
the nonterminal $S$ by repeated application of the production rules. We write
$A \stackrel{\bullet}{\to} w$ for a tuple of strings $w\in (\fA^*)^{d(A)}$ that is obtained 
from a nonterminal $A\in \cN$ by repeated application of the production rules of $\cG$. 
So $\cL_\cG=\{ w\in \fA^* \,|\, S \stackrel{\bullet}{\to} w \}$.
\end{defn}

\smallskip

The following example (from \cite{ClaYo}) illustrates the general definition above.

\begin{ex}\label{MCFGex} {\rm
As an example, the $2$-MCFG $\cG$ with $\fA=\{ a,b,c,d \}$ and $\cN=\{ S, N, M \}$ with $d(N)=d(M)=2$ and
production rules
$$ S\mapsto f(N,M), \ \ \  N \mapsto (a,c), \ \ \  M \mapsto (b,d), \ \ \  N \mapsto g(N), \ \ \  M \mapsto h(M) \, , $$
with $f((x_1,x_2), (y_1,y_2))=x_1y_1x_2y_2$, $g(x_1,x_2)=(x_1 a, x_2 c)$, and $h(y_1,y_2)=(y_1 b, y_2 d)$,
computes the language $\cL_\cG=\{ a^n b^m c^n d^m \,|\, n,m\in \N \}$. This example contains a cross-serial
dependence that is not realizable by a context-free grammar, showing the context-sensitivity of MCFGs. 
}\end{ex}

\smallskip

\begin{rem}\label{FSSigmaS}{\rm
Given an m-MCFG $\cG=(\fA,\cN,P,S)$ as in Definition~\ref{MCFGdef}, we denote by 
$F_\cG$ the set of functions $f(B_1,\ldots, B_q)$ associated to production rules of the
form $S\mapsto f(B_1,\ldots, B_q)$ and by $\Sigma_\cG$ the set of all tuples of strings 
$(\omega_1, \ldots, \omega_q)$ where $\omega_k\in (\fA^*)^{d(B_k)}$ satisfies 
$B_k \stackrel{\bullet}{\to} \omega_k$ 
for $(B_1,\ldots, B_q)$ the argument of some $f\in F_\cG$. Thus the m-MCF language $\cL$
is the image $F_\cG(\Sigma_\cG)$, where each $f(B_1,\ldots, B_q)\in F_\cG$ applies to a 
domain in $\Sigma_\cG$ consisting of the tuples $(\omega_1, \ldots, \omega_q)$ with 
$B_k \stackrel{\bullet}{\to} \omega_k$. 
Thus, it is equivalent to specify $\cL$ as $\cL=\{ w\in \fA^* \,|\, S \stackrel{\bullet}{\to} w \}$ or
by specifying $\Sigma_\cG$ so that $\cL=F_\cG(\Sigma_\cG)$. The latter description in terms of the
set of tuples $\Sigma_S$ is preferable for our purposes, as we will see below.  Note that while
the language $\cL$ may be realizable by different grammars $\cG$, the sets $\Sigma_\cG$ and
$F_\cG$ explicitly depend on the production rules of the given grammar $\cG$ so they depend
on $\cG$, not just on $\cL$.
}\end{rem}

\smallskip

We can extend the category of context-free languages constructed in \S \ref{FLcatSec} to a
category of multiple context-free grammars. (Here we need to work with the richer data of the
grammars rather than simply with the resulting languages because of the observation in
Remark~\ref{FSSigmaS}.)

\smallskip

\begin{defn}\label{Mtransd}
A matrix rational transduction $X=(X_{k\ell})_{k,\ell=1}^N$ consists of an $N\times N$
matrix where each $X_{k\ell}$ is a rational transduction as in Definition~\ref{rattransd}.
\end{defn}

\smallskip

The following construction of a category of multiple context-free grammars follows 
directly from the construction for the context-free case with the same argument
adapted to the matrix form of morphisms.

\begin{lem}\label{MCFGcat}
Let $\Sigma_\cG$ and $F_\cG$ be as in Remark~\ref{FSSigmaS} and let $\Sigma_{\cG,k}$ be 
the projection of tuples $(\omega_1, \ldots, \omega_q)\in \Sigma_\cG$ onto
the $k$-th component $\omega_k \in (\fA^*)^{d(B_k)}$.
Multiple context-free grammars form a category $\cM\cC\cF\cG_m$, where the set of objects
${\rm Obj}(\cM\cC\cF\cG_m)$ consists of the m-MCFGs $\cG=(\fA,\cN,P,S)$ as in Definition~\ref{MCFGdef},
and the set of morphisms ${\rm Mor}_{\cM\cC\cF\cG_m}(\cG,\cG')$ consists of 
matrix transductions $X=(X_{k\ell})_{k,\ell=1}^q$ where 
each $X_{k\ell}$ is a matrix transduction between $X_{k\ell}: (\fA^*)^{d(B_k)} \to (\fB^*)^{d(B'_\ell)}$,
such that $X_{k\ell}(\Sigma_{\cG,k})\subset \Sigma_{\cG',\ell}$ and such that, for any $f' \in F_{\cG'}$
the precomposition $f=f'\circ X$ is in $F_{\cG}$, and such that all the $X_{k,\ell}$ have the same $\cL_{reg}$. 
\end{lem}

\medskip

\subsection{Spin systems from multiple context-free grammars} \label{spinsMCFGsec}

We extend the category of spin chains of \S \ref{spincatSec} 
to a category of $k$-spin systems in the following way. 

\begin{defn}\label{kspinsys}
Let $\Upsilon^{(k)}=(\fA, \Omega^{(k)}, \underline{J}_{\fA^k})$ be as in Definition~\ref{SpinSetDef}
but with $\Omega^{(m)}\subset (\fA^*)^k$ a non-empty set of $k$-tuples of words
(where some but not all of the coordinates may be equal to the empty word $\epsilon$).  
A spin system $S_{\Upsilon^{(k)}}$ is given by 
\begin{equation}\label{mSsys}
S_{\Upsilon^{(k)}} = \bigsqcup_{\omega =(\omega_1,\ldots, \omega_k) \in \Omega^{(k)}} S_\omega\, ,
\end{equation}
where $S_\omega$ is a spin system with interacting spins 
$\sigma_{a_{1,i_1},\ldots, a_{k,i_k}, \epsilon_1,\ldots, \epsilon_k}$, for 
$\omega_i=a_{i_0}\cdots a_{i_{\ell(\omega_i)}}$ and with $\epsilon_i=\{ \pm 1 \}$,
located at the $2^k$ vertices of a $k$-dimensional cube with center labelled by
$(a_{1,i_1},\ldots, a_{k,i_k})$. These satisfy the identifications
$$\sigma_{a_{1,i_1},\ldots, a_{i,j}, \ldots, a_{k,i_k}, \epsilon_1,\ldots,\underbrace{ + }_{\text{i-th}},
 \ldots \epsilon_k}= \sigma_{a_{1,i_1},\ldots, a_{i,j+1}, \ldots, a_{k,i_k}, \epsilon_1,\ldots, 
 \underbrace{ - }_{\text{i-th}} , \ldots \epsilon_k}\ , . $$
The interaction is described by Boltzmann weights
for spins on the vertices of a $k$-cube of the form
\begin{equation}\label{BoltzWm}
W^{(k)}(\sigma_{\underline{\epsilon}}) = W(\sigma_{\underline{\epsilon}(0)}| 
\sigma_{\underline{\epsilon}(1)} | \cdots | \sigma_{\underline{\epsilon}(k-1)} | \sigma_{\underline{\epsilon}(k)}) \, , 
\end{equation}
for $\underline{\epsilon}\in \{ \pm 1 \}^k$ and with $\underline{\epsilon}(j)$ denoting the
vectors in $\{ \pm 1 \}^k$ with $j$ entries equal to $-1$, and with the notation
$\sigma_{A}:=(\sigma_a)_{a\in A}$ for a subset $A\subset \{ \pm 1 \}^k$. The set of vectors
$\underline{J}_{\fA^k}$ consists of all the couplings involved in the interaction terms in the
Boltzmann weight \eqref{BoltzWm}. 
\end{defn}

Here we do not make any more specific assumption on the form of the 
Boltzmann weight \eqref{BoltzWm}. Its dependence on a set of spins on
the vertices of a hypercube places these spin systems in the category of
vertex models, and the form \eqref{BoltzWm} of the Boltzmann weight
is designed to generalize the case of the integrable Zamolodchikov model
in 3D, to which we will return below. We assume here only that the 
Boltzmann weight \eqref{BoltzWm} depends on a set of real parameters
that specify the couplings of the interaction terms between the spins
$\sigma_{a_{1,i_1},\ldots, a_{k,i_k}, \epsilon_1,\ldots, \epsilon_k}$.
As in the case of the aperiodic spin chains discussed above, these 
parameters are in general not assumed to be constant but depend
on the location in the spin system through a dependence on
the position $(a_{1,i_1},\ldots, a_{k,i_k})$ of the spins hence 
we describe these parameters as vectors $\underline{J}_{a_1,\ldots, a_k}\in 
\underline{J}_{\fA^k}$. 

\begin{defn}\label{sysRenDef}
A spin system with sets of spins assigned to location defined by
words $\omega\in (\fB^*)^k$ and with interaction couplings depending 
on parameters $\underline{J}_{a_1,\ldots, a_k}\in 
\underline{J}_{\fB^k}$ admits a block spin renormalization if
there is an injective homomorphism $\alpha: \fA^* \to \fB^*$ of free monoids
and a new set of parameters $\underline{\hat J}_{\fA^k}$ such that
$$ W^{(k)}_{\underline{J}_{\fB^k}}(\sigma_{\omega=\alpha(\eta)}) 
=W^{(k)}_{\underline{\hat J}_{\fA^k}}(\sigma_{\eta})\, . $$
We write $$\underline{\hat J}_{\fA^k}=\cR_\alpha(\underline{J}_{\fB^k})\, . $$
\end{defn}

In the case of a homomorphism $\alpha: \fA^* \to \fB^*$ that has
a nontrivial kernel, the notion of block renormalization (spin decimation)
described in Definition~\ref{sysRenDef} needs to be replaced
by a combination of decimation and disconnection as discussed
earlier, the result of which one can still denote by 
$\underline{\hat J}_{\fA^k}=\cR_\alpha(\underline{J}_{\fB^k})$.

\smallskip

Let $\Pi_i$ with $i\in \{ 1, \ldots, k \}$ denote the projection $\Pi_i : (\fA^*)^k \to \fA^*$
onto the $i$-th coordinate and let $\Pi_i(S_{\Upsilon^{(k)}})$ be the spin chain with spins
\begin{equation}\label{iprojPi}
\Pi_i(S_{\Upsilon^{(k)}})=\sqcup_{\omega_i \in \Pi_i(\Omega^{(k)})} S_{\omega_i} \, ,
\end{equation}
with $S_{\omega_i}$ consisting of spins $\sigma_{a_{i,j}, \pm}$ with 
$\sigma_{a_{i,j},+}=\sigma_{a_{i,j+1},-}$ and with interaction term
obtained from the interaction terms of \eqref{BoltzWm} by
summing out the other spins.  

\smallskip

We can then form a category of such $k$-spin systems that generalizes the
category of spin chains as the $k=1$ case. Again this follows the same
argument given for spin chains. 

\begin{lem}\label{kScat}
Given two $k$-spin systems $S_{\Upsilon^{(k)}}$ and $S_{\Upsilon^{(k)} \prime}$,
with $\Upsilon^{(k)}=(\fB, \Omega^{(k)}, \underline{J}_{\fB^k})$ and
$\Upsilon^{(k) \prime}=(\fC, \Omega^{(k) \prime}, \underline{J}'_{\fC^k})$, 
a morphism $\bD(S_{\Upsilon^{(k)} \prime},S_{\Upsilon^{(k)}})$ consists of a
triple $(S_{\hat \Upsilon^{(k\times k)}}, \underline{\alpha}, \underline{\beta})$ with a 
$k$-spin system 
$$ S_{\hat \Upsilon^{(k\times k)}} =\sqcup_{\eta=(\eta_i)_i \in \hat\Omega^{(k\times k)}} S_{\eta_i} $$
$$ \hat \Upsilon^{(k\times k)} = (\fA_o, \hat\Omega^{(k\times k)} , \underline{\hat J}_{\fA_o^k}) $$
$$ \hat\Omega^{(k\times k)}= (\underline{\alpha}\times \underline{\beta})(\cL_{reg}^k) \cap (\Omega^{(k)}\times
\Omega^{(k) \prime}) $$
with homomorphisms $\underline{\alpha}=(\alpha_{i,j}): (\fA^*)^k \to (\fB^*)^k$ and $\underline{\beta}=(\beta_{ij}) : (\fA^*)^k \to (\fC^*)^k$, with $\alpha_{ij}: \Pi_i (\fA^*)^k \to \Pi_j (\fB^*)^k$ and
$\beta_{ij}: \Pi_i (\fA^*)^k \to \Pi_j (\fC^*)^k$. The vector $\underline{\hat J}_{\fA^k}$ is 
determined by $\underline{J}_{\fB^k}$ and $\underline{J}_{\fC^k}$ with 
$$ \underline{\hat J}_{\fA^k}=\cR_{\underline{\alpha}}(\underline{J}_{\fB^k})=
\cR_{\underline{\beta}}(\underline{J'}_{\fC^k})\, ,  $$
namely the compatibility
condition $$ \underline{\hat J}_{a=(a_1,\ldots, a_k)}=\cR( \underline{J}_{\underline{\alpha}(a)} )
=\cR( \underline{J}'_{\underline{\beta}(a)} ) \, . $$
\end{lem}

We then obtain, as in the context-free case, a way of assigning spin
systems to formal languages compatible with this categorical structure.

\begin{prop}\label{MCFGfunctor}
Let $\cM\cC\cF\cG_m\cJ$ denote the category with objects given by pairs $(\cG, J)$ with
$\cG\in {\rm Obj}(\cM\cC\cF\cG_m)$ and $J: \fA^m \to \R^{m \cdot N\cdot \cdot \# \fA}$, for
some fixed $N\in \N$, with $\underline{J}_{\fA^m}:=J(\fA^m)$. Morphisms $(X,\hat J)$ in
${\rm Mor}_{\cM\cC\cF\cG_m\cJ}((\cG,J),(\cG',J'))$ are morphisms $X$ in 
${\rm Mor}_{\cM\cC\cF\cG_m}(\cG,\cG')$ with $\hat J: \fA^m \to 
\R^{m \cdot N\cdot \cdot \# \fA}$ with $\hat J=\cR(J)=\cR(J')$. The construction of
spin systems associated to multiple context-free grammars is the functorial
assignment determined by $\fS(\cG,J)=S_{\Upsilon^{(m)}}$ with $\cG=(\fA, \cN, P, S)$
with corresponding sets $\Sigma_\cG$ and $F_\cG$ as above, and 
$\Upsilon^{(m)}=(\fA,\Sigma_\cG,\underline{J}_{\fA^m})$ and morphisms 
$\fS(X,\hat J)=(\bD(\Upsilon^{(m) \prime},\Upsilon^{(m)}), \hat J)$, with
$\bD(\Upsilon^{(m) \prime},\Upsilon^{(m)})$ as in Lemma~\ref{kScat}
with $\hat\Omega^{(m\times m)}=X \cap (\Sigma_\cG \times \Sigma'_{\cG'})$. 
\end{prop}

Here $N\in \N$ will correspond the number of parameters (the length of the vector $\underline{J}$ that
are needed to specify the interaction terms of the Boltzmann weights in the construction of the
associated spin system.

\section{The icosahedral quasicrystal}\label{IcosaqlattSec}

As one of our main motivating example, we consider the construction of exactly solvable
spin models on quasicrystals introduced in \cite{Ko}. While the discussion in \cite{Ko}
also covers the case of the 2D Penrose tilings, we focus here on the case of the
3D icosahedral quasicrystal, which we show is a case of the general construction 
of spin systems from MCFGs that we developed in the previous sections. We first
give an overview of the geometry of this model, in \S \ref{ABCKsec} by comparing
different descriptions of the icosahedral quasicrystal, with a group-theoretic proof 
of their equivalence. We then focus on the description
in terms of the Ammann planes. 
We construct in \S \ref{AmmannFLsec} a MCFG that describes the quasicrystal geometry
encoding the icosahedral Ammann quasilattice, with the D0L substitution
rules as an endomorphism. We start the construction of the
spin model associated to this MCFG 
in \S \ref{AmmannSec2} where we show how the tuples of words in $\Sigma_\cG$ for the 
MCFG of  \S \ref{AmmannFLsec} map to regions of the  icosahedral Ammann quasilattice
under the mapping of Proposition~\ref{MCFGfunctor}. We then show in \S \ref{IcoModelSec} 
that the associated spin model is the icosahedral quasicrystal integrable spin model, by
showing that it has the expected Botlzmann weights, with the datum $J$ of the
parametrs of the model given by the spectral parameters of \cite{Bax2}. 

We recall two further descriptions of the Boltzmann weights: one in \S \ref{IcoModelSec},
based on \cite{Bax3}, in terms of the transfer matrix and 
body-centered-cube (BCC) lattice. This allows us to show that the assignment
of the datum $J$ in our functorial construction of spin systems from MCFGs
can in general be done in different possible way, meaning that different
assignments of the datum $J$ in the objects of the category 
$\cM\cC\cF\cG\cJ$ of Proposition~\ref{MCFGfunctor} can 
correspond to the same spin model. We also recall in \S \ref{solvSec}, based on
\cite{Bax2}, the reformulation of the Boltzmann weights in terms of Zamolodchikov spins, 
which is used to show solvability and the relation to the Zamolodchikov
 tetrahedron equation.
 
 In \S \ref{solvSec} we return to the general setting of Proposition~\ref{MCFGfunctor} 
 and we prove that the tetrahedron equation for the Boltzmann weights of the
 icosahedral quasicrystal spin model have a natural generalization for the  
 Boltzmann weights on higher dimensional hypercubes of \eqref{BoltzWm},
 obtained in terms of two dual cubulations of the $n$-simplex. 

Finally, we discuss some additional properties of the icosahedral quasicrystal spin
model, the bulk free energy computation based on \cite{Bax} and \cite{Bax3}, and
the criticality property, following the argument of \cite{Bax}. 
We suggest as final questions the possibility of extending these types of arguments for
other spin systems associated to formal languages as in Proposition~\ref{IAQmCFG}.

\smallskip

\subsection{Danzer’s ABCK tiling and the Socolar-Steinhardt tiling}\label{ABCKsec}

\begin{figure}[h]
    \centering
    \includegraphics[scale=0.4]{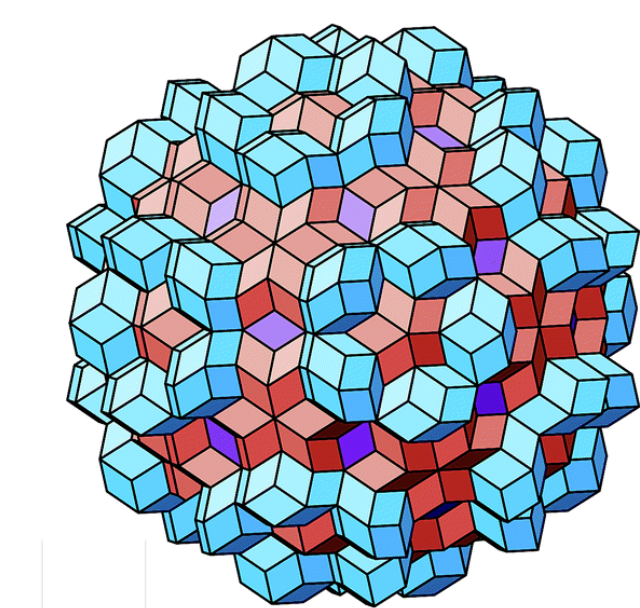}
    \caption{Icosahedral packing with rhombic triacontahedron, rhombic 
    dodecahedron, rhombic icosahedron, and prolate rhombohedra prototiles \cite{Mad}.  \label{FigSocSt} }
\end{figure}

There are three widely-used constructions for the icosahedral packing (also referred to as tilings
by extension of the terminology adopted in the 2D case), 
with the following prototiles:
\begin{enumerate}
    \item Acute (or prolate) and obtuse (or oblate) golden rhombohedra \label{rhomb_til}
    \item Rhombic triacontahedron, rhombic dodecahedron, rhombic icosahedron, and prolate rhombohedra \cite{SocStein} \label{soc_stein_til}
    \item Danzer's $A, B, C, K$ tetrahedra \cite{Danzer} \label{danz_til}
\end{enumerate}
The first construction is known as the Ammann rhombohedral tiling (\ref{rhomb_til}), the second (\ref{soc_stein_til}) as the Socolar-Steinhardt tiling (displayed in Figure~\ref{FigSocSt}) 
\cite{SocStein}, and the third as Danzer's $ABCK$ tiling. 
Icosahedral packings in particular are often referred to as Penrose tilings (in 3D).

\begin{figure}[h]
    \centering
    \includegraphics[scale=0.5]{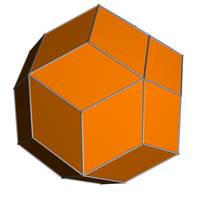}
    \caption{One of the four prototiles of the Socolar-Steinhardt tiling, the rhombic triacontahedron $T$, 
    can be decomposed into prolate and oblate golden rhombohedra. \label{FigRhombic}}
\end{figure}

These three constructions are known to be equivalent, see \cite{Roth}, \cite{Danzer}. 
We show here in a direct explicit way how to construct one of these icosahedral packings from the other,
for instance, the Socolar--Steinhardt tiling from the Danzer $ABCK$ tiling. This 
involves some geometric manipulation of the prototiles for each structure. In this section
we focus on the Socolar--Steinhardt tiling and Danzer's $ABCK$ tiling. 
The inflation rules for these tilings are context-free, whereas, as we will discuss later, 
the inflation rules for the Ammann rhombohedral tiling are context-sensitive, see
\cite{SocStein} and \cite{Mad} for further discussion of context-free versus context-sensitive inflation rules. 
The Socolar-Steinhardt tiling and Danzer's $ABCK$ tiling are also of particular interest because they 
are face-centered icosahedral (FCI) quasilattices, corresponding to the stable AlFeCu FCI quasicrystals 
(see \cite{Roth}, \cite{Cornier}). To illustrate the explicit relation between these
tiling constructions we show how to obtain the Socolar-Steinhardt tiling 
using Danzer's $ABCK$ tiling, that is, we show the construction of the 
rhombic triacontahedron (Figure~\ref{FigRhombic}) using Danzer's tetrahedra. 
We then note that the rhombic dodecahedron and rhombic icosahedron can 
be constructed using rhombic triacontahedron, and the rhombic triacontahedron 
can be decomposed into prolate and oblate golden rhombohedra, so that 
(\ref{danz_til}) can be used to construct both (\ref{rhomb_til}) and (\ref{soc_stein_til}). 
We give an explicit group-theoretic proof of the equivalence of these three structures,
which is simpler than other proofs based on algebraic topology or local matching rules,
\cite{Roth}, \cite{Danzer}: we first show that Danzer's $K$ tetrahedra form the fundamental 
region from which $T$ can be constructed, and we give the 
explicit construction of $T$ using $K$. 

\smallskip
\subsubsection{Coxeter groups}

We review some terminology on Coxeter groups and root systems. For a more detailed 
review, see for instance \cite{Kane}. We use the notation $(x,y) = \Vert x \Vert \Vert y \Vert \cos\theta$ 
for the angle $0\leq\theta\leq\pi$ between points $x, y$ in a vector space $V$. 

\begin{itemize}
\item  Given a hyperplane $P$ in a vector space $V$, define the {\em reflection} with respect to $P$ as the linear transformation $S: V \rightarrow V$ such that $Sx = x$ for $x \in P$ and $Sx = -x$ for $x \in P^{\perp}$. If $P^\perp = \mathbb{R}r,$ then $Sx = x - 2\frac{(x,r)}{(r,r)}r$.
    \item Let $V$ be a vector space, and $T$ a linear transformation. Then the {\em orthogonal group} $O(V)$ is the set of all linear transformations $T$ of $V$ which preserves angles between vectors, such that $(Tx, Ty) = (x, y)$.
\item A subgroup $G$ within $O(V)$ is {\em effective} if the elements of $G$ share no common fixed points; for all $T \in G$ and points $x \in V, Tx \neq x$. 
\item Let $O(V)$ be an orthogonal group. Then $G$ is a {\em Coxeter group} if it is a finite, effective subgroup of $O(V)$ which is generated by reflections.
\item     If $S \in G$ is a reflection through the hyperplane $P = r^\perp$, and $r$ has length one, then $r$ and $-r$ are the {\em roots} of $S$, and the roots of $G$ are the roots of the reflections in $G$.
 \item        Fix a set of generators consisting of reflections for $G$. Let $\Delta$ be the set consisting of the union of orbits of these roots under the action of $G$. Since $\Delta$ is the union of orbits, $G$ acts on $\Delta$ and $\Delta$ is a {\em root system} for $G$.
    \item     A Coxeter group $G$ is {\em crystallographic} if there exists a lattice $\mathbb{R}^n$ which is stabilized by $G$.
    \item Given a root system $\theta$, its {\em rank} $d$ is the dimension of the vector space over $\mathbb{R}$ generated by taking all \textit{real} linear combinations of the roots, and its {\em rational rank} $d_\mathbb{Q}$ is the dimension of the vector space over $\mathbb{Q}$ generated by taking all \textit{rational} linear combinations of the roots. 
\end{itemize}
Crystallographic root systems come in four infinite families
 \{$A_n (n \geq 1), B_n (n \geq 2), C_n (n \geq 3), D_n (n \geq 4)$\} 
 and five exceptional cases ($G_2, F_4, E_6, E_7, E_8$). All other root systems are
  noncrystallographic; most of them occur in 2D ($I_2^n, n = 5, 7, 8, 9,...$), with one 
  occurring in 3D ($H_3$) and one in 4D ($H_4$). The $H_3$ case is of particular 
  interest to us here.

Let $\tau$ be the golden ratio and $\sigma$ its Galois conjugate:
\begin{equation} \label{const1}
    \begin{gathered}
        \tau = \frac{1+\sqrt{5}}{2}, \ \ \ \ \ \ 
        \sigma = -\tau^{-1} = \frac{1 - \sqrt{5}}{2}
    \end{gathered}
\end{equation}
The roots of $H_3$ are $30$ vectors obtained from
\begin{equation} \label{H3_roots}
    \begin{gathered}
        \{\pm1, 0,0\}, \ \ \ \ \ 
        \frac{1}{2}\{\pm\tau, \pm1,\pm\sigma\}
    \end{gathered}
\end{equation}
by taking all combinations of $\pm$ signs and all even permutations of the three coordinates. 
These roots point to the $30$ edge midpoints of a regular icosahedron \cite{Cox} and the corresponding reflections generate the full symmetry group of the icosahedron, which is of order $120$.
When $\theta$ is crystallographic, $d_\mathbb{Q} = d$, but when $\theta$ is noncrystallographic, 
$d_\mathbb{Q} > d$. So in the noncrystallographic case, it seems as if there are two dimensions for the root system. The roots of $\theta$ live in $\mathbb{R}^d$ in one sense, but they also live in the higher-dimensional space $\mathbb{Q}^{d_\mathbb{Q}}$ in another sense. 
For example, the Coxeter group $H_3$ has an ordinary rank $d = 3$ but a rational rank $d_\mathbb{Q} = 6$.

We review the notion of a Coxeter pair (see also \cite{BoStein}).
Consider two irreducible root systems $\theta$ and $\theta^{\parallel}$. $\theta$ is a crystallographic root system of rank $d$, whose $j$th root, denoted $\textbf{r}_j$, corresponds to a reflection $R_j$ that acts on the $d$-dimensional coordinates $\textbf{x}$ as 
\begin{equation} \label{theta_ref}
    R_j: \textbf{x} \rightarrow \textbf{x} - 2\frac{(\textbf{x},\textbf{r}_j)}{(\textbf{r}_j,\textbf{r}_j)}\textbf{r}_j.
\end{equation}
$\theta^\parallel$ is a noncrystallographic root system of (ordinary) rank $d^\parallel$, whose $j$th root, denoted $\textbf{r}_j^\parallel$, corresponds to a reflection $R_j^\parallel$ that acts on the $d^\parallel$-dimensional coordinate $\textbf{x}^\parallel$ as 
\begin{equation} \label{theta_par_ref}
    R_j^\parallel: \textbf{x}^\parallel \rightarrow \textbf{x}^\parallel - 2\frac{(\textbf{x}^\parallel,\textbf{r}_j^\parallel)}{(\textbf{r}_j^\parallel,\textbf{r}_j^\parallel)}\textbf{r}_j^\parallel.
\end{equation}

\begin{defn}\label{Coxpair}
$\theta$ and $\theta^\parallel$ form a {\em Coxeter pair} of degree $N = \frac{d}{d^\parallel}$ if 
\begin{enumerate}
    \item they have the same rational rank
    \item $N$ copies of the $\theta^\parallel$ roots are obtained from the maximally symmetric orthogonal projection of the $\theta$ roots onto a $d^\parallel$-dimensional plane
\end{enumerate}
\end{defn}

If $d^\parallel = 3$, then $\theta^\parallel = H_3$ with rational rank $d_\mathbb{Q} = 6$. Its partner crystallographic root system (with rational rank $d = 6$) is $D_6$ because it is the only rank-six root system whose total number of roots (60) is a multiple of the total number of $H_3$ roots (30). Furthermore, $D_6$ has a projection onto $d^\parallel = 3$ dimensions with $G(H_3)$ (icosahedral) symmetry, where the projected roots split into two copies of $H_3$. The Coxeter pair $(H_3, D_6)$ has degree $N = 2$. See Appendix A of \cite{BoStein} for more details. 

Recall that a fundamental domain or fundamental region of the action of a group on a set 
is a set of representatives for the orbits of the action. 

\begin{prop}\label{DanzerF}
Danzer's $K$ tetrahedron forms the fundamental 
region from which the rhombic triacontahedron $T$ can be generated. 
\end{prop}

\proof Let $a,b$ be given by
\begin{equation} \label{const2}
        a = \frac{\sqrt{10+2\sqrt{5}}}{4},  \ \ \ \ \ \ 
        b = \frac{\sqrt{3}}{2} 
\end{equation}
Consider the Coxeter pair $(H_3, D_6)$ as above. One obtains the icosahedral packing 
by projecting the three-dimensional boundaries of the cells of the $D_6$ lattice, \cite{Kramer}. 
A vector of the $D_6$ lattice can be written as a linear combination of the roots 
with integer coefficients:
\begin{equation} \label{D_6_vectors}
    \lambda \sum_{i=1}^{6} m_il_i,
\end{equation}
where $m_i \in \mathbb{Z}$ and $\{l_i\}$ is an orthonormal set of vectors. We follow the 
construction in \cite{alSiy} to define the weights of $D_6$ and $H_3$. The weights of $D_6$ are 
\begin{equation} \label{D_6_weights}
        w_f = \sum_{j=1}^{f}l_j, \,\,\, f \in \{1,2,3,4\}, \ \ \ \ \
        w_5 = \frac{1}{2}(-l_6 + \sum_{j=1}^{5}l_j), \ \ \ \ \ 
        w_6 = \frac{1}{2}\sum_{j=1}^{6}l_j . 
\end{equation}
Define the roots and weights of $H_3$ in two complementary three-dimensional spaces $E_\perp$ and $E_\parallel$, respectively. The weights of $H_3$ in $E_\parallel$ are given by 
 $$       v_1 = \frac{1}{\sqrt{2+\tau}}(w_1 + \tau w_5) = \frac{1}{\sqrt{2}}(1,\tau,0),  \ \ \ \ \ 
        v_2 = \frac{1}{\sqrt{2+\tau}}(w_2+\tau w_4) = \sqrt{2}(0,\tau,0),  $$
   $$     v_3 = \frac{1}{\sqrt{2+\tau}}(w_6+\tau w_3) = \frac{1}{\sqrt{2}}(0,\tau^2,1),  $$
and they denote the five-fold, two-fold, and three-fold symmetry axes of the icosahedral group, respectively. By applying the group elements on the weight vectors, we obtain the orbits $v_1$ and $v_3$. 
(The orbit of $v_2$ is not relevant here):
\begin{equation} \label{orbits}
\begin{array}{ll}
        \frac{(v_1)_{H_3}}{\sqrt{2}} = & \frac{1}{2}\{(\pm1, \pm\tau, 0), (\pm\tau, 0, \pm1), 
        (0, \pm1, \pm\tau)\},  \\[2mm] 
        \frac{(\tau^{-1}v_3)_{H_3}}{\sqrt{2}} =&  \frac{1}{2}\{(\pm1, \pm1, \pm1), (0, \pm\tau, \pm\sigma), (\pm\tau, \pm\sigma, 0), (\pm\sigma, 0, \pm\tau)\}
        \end{array}
\end{equation}
The notation $(v_i)_{H_3}$ is introduced for the set of vectors generated by the action of the icosahedral group on the weight $v_i$. The orbits in \eqref{orbits} represent the vertices of an icosahedron and dodecahedron, respectively, in the $E_\parallel$ space. The union of the orbits of \eqref{orbits} constitutes the vertices of the rhombic triacontahedron $T$. Lastly, the generating relations of $H_3$ are given by 
$$  R^2_1 = R^2_2 = R^2_3  = (R_1R_3)^2 = (R_1R_2)^3 = (R_2R_3)^5 = 1 $$
\begin{equation} \label{generating_rel}
        R_1 = \begin{bmatrix}
                -1 & 0 & 0\\
                0 & 1 & 0\\
                0 & 0 & 1
            \end{bmatrix},  \ \ \ \ \ 
        R_2 = \frac{1}{2}\begin{bmatrix}
            1&-\sigma&-\tau \\ -\sigma&\tau&1 \\
            -\tau&1&\sigma
        \end{bmatrix}, \medskip  \ \ \ \ \  
        R_3 = \begin{bmatrix}
                1&0&0\\
                0&1&0\\
                0&0&-1
            \end{bmatrix} \, .
\end{equation}
The Danzer $K$ tetrahedron is defined in \cite{Danzer} by specifying the dihedral angle 
and the length of the corresponding edge for each pair of vertices:
\begin{equation}
    \{(\frac{\pi}{2}, \frac{\tau}{2}), (\frac{\pi}{2}, \frac{\tau^{-1}}{2}), (\frac{\pi}{2}, \frac{1}{2}), (\frac{\pi}{3}, b), (\frac{\pi}{5}, a), (\frac{2\pi}{5}, \tau^{-1}a)\}
\end{equation}
The coordinates of the four vertices are given by
\begin{equation}
        D_{0} = (0,0,0), \ \ \ \ 
        D_{1} = \frac{v_1}{\sqrt{2}} , \ \ \ \ \
        D_{2} = \frac{v_2}{2\sqrt{2}}, \ \ \ \ \
        D_{3} = \frac{\tau^{-1}v_3}{\sqrt{2}},
\end{equation}
where $D_0$ is the origin. 

The union of the orbits of \eqref{orbits} constitutes the set of
vertices of $T$. One of $T$'s cells is a pyramid based on a 
golden rhombus with vertices 
\begin{equation} \label{pyr_vertices}
    \frac{v_1}{\sqrt{2}},\ \ \ \  \frac{\tau^{-1}v_3}{\sqrt{2}}, \ \ \ \ R_1\frac{v_1}{\sqrt{2}}, \ \ \ \ R_3\frac{\tau^{-1}v_3}{\sqrt{2}}, \ \ \ \  (0,0,0),
\end{equation}
where the apex is at the origin.
 Furthermore, the intersection of the diagonals of the rhombus occurs at 
 $\frac{v_2}{2\sqrt{2}}$ and its magnitude is the in-radius of $T$. 
 Therefore, the weights $\frac{v_1}{\sqrt{2}}, \frac{\tau^{-1}v_3}{\sqrt{2}}, \frac{v_2}{2\sqrt{2}}$ 
 can be taken as the vertices of $K$. Thus, we have shown that $K$ is the fundamental 
 region of the icosahedral group that generates $T$.
\endproof

One can see the construction of $T$ from $K$ explicitly by first using the tetrahedron 
$K$ to construct an octahedron $\tilde{K}$, which contains the pyramid cell forming 
the rhombic triacontahedron $T$. Define $K' = R_1K$ as the mirror image of $K$  
and note that $T$ consists of $60K + 60K' = 30(2K + 2K')$, where $2K + 2K'$ forms 
the pyramid cell of $T$ described earlier. This cell has edge length $\tau^{-1}a$ and 
height $\frac{\tau}{2}$. To construct $2K+2K'$, we introduce the octahedron $\tilde{K}$, which consists of $8K$ generated by a group of order eight (Figure 4). The group of order eight consists of three commuting generators: $R_1, R_3, R_0$. $R_1$ and $R_3$ are as defined above, and $R_0$ is an affine reflection with respect to the golden rhombic face induced by the affine Coxeter group $\tilde{D}_6$. 
The octahedron $\tilde{K}$ can be dissected into three non-equivalent pyramids with rhombic 
bases by cutting along the lines orthogonal to the three planes $v_1-v_2, v_2-v_3,v_3-v_1$. 
One of these pyramids is the pyramid cell of $T$. When rotated by the icosahedral group, 
this pyramid generates $T$. The octahedron $\tilde{K}$ includes the vertices of $2K+2K'$ 
in \eqref{pyr_vertices}, as well as $(0,0,0)$ and $(0,\tau,0)$. A mirror image of $4K = 2K+2K'$ 
with respect to the rhombic plane, corresponding to an affine reflection, complements it up 
to the octahedron $\tilde{K}$.
After generating $T$, one can use $T$ to construct the rhombic dodecahedra 
and icosahedra of the Socolar-Steinhardt tiling. To obtain the prolate rhombohedra, 
one can decompose $T$ into prolate and oblate rhombohedra. This also enables 
one to obtain the Ammann rhombohedral tiling. 

Equivalence between aperiodic packings is useful, for instance, for calculating frequencies 
of certain patterns or vertex configurations. In the case of the Danzer and Socolar-Steinhardt tilings, 
deriving the vertex configurations for the latter is quite complicated, but for the former quite simple.

\subsection{Formal language for the icosahedral Ammann quasilattice}\label{AmmannFLsec}

\begin{figure}
    \begin{center}
    \includegraphics[scale=0.5]{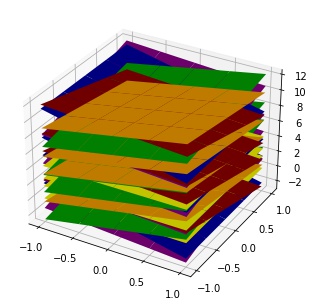}
    \caption{Icosahedral Ammann quasilattice corresponding to the icosahedral quasicrystal. Each set of colored planes is orthogonal to an icosahedral star vector $\Vec{e}_j$. Planes of one set alternate between $\bL$ (long) and $\bS$ (short) spacing. The axes are arbitrary, depending on how many inflations the quasilattice has undergone.}
    \label{fig:IAQ}
    \end{center}
\end{figure}

Packings for quasicrystals are comprised of polyhedra prototiles, and can be 
constructed using inflation and deflation rules, \cite{Mad}. An inflation rule 
replaces every prototile in the packing with other prototiles arranged in a 
specific orientation. This procedure can be repeated infinitely many times to
construct an aperiodic structure that extends indefinitely. 
The icosahedral quasicrystal, as we discussed above, consists of rhombic triacontahedra, 
rhombic dodecahedra, rhombic icosahedra, and prolate rhombohedra as prototiles. 
When inflating the icosahedral quasicrystal, each rhombic dodecahedron, for example, 
is replaced by a combination of rhombic dodecahedra, rhombic triacontahedra, rhombic 
icosahedra, and prolate rhombohedra arranged in a specific formation \cite{Mad}. 
Due to the complexity of these prototiles and the inflation rules, this is not the simplest
way to describe the geometric structure of the quasicrystal. In fact, packings are sometimes
better described using a dual transformation,  \cite{BoStein}, \cite{SocStein}. 
We refer to a quasicrystal's dual structure as its {\em Ammann quasilattice}. 
Figure~\ref{fig:IAQ} displays the icosahedral Ammann quasilattice (IAQ) dual  
to the icosahedral quasicrystal. 

The IAQ consists of six sets of planes $k_j$, 
where $j$ ranges from zero to five. Each set of planes is orthogonal to an 
icosahedral star vector $\Vec{e}_j$ (represented by a different shade/color 
in Figure ~\ref{fig:IAQ}), see \cite{SocStein}. Between planes of one set, 
the spacing alternates between $\bS$ (short) and $\bL$ (long), where 
$\bL/\bS = \tau$, the golden ratio. The planes of the IAQ divide the dual 
space into polyhedra. Define the polygons comprising each polyhedra as the {\em faces} 
of the quasilattice.

We now construct a formal language encoding the geometric structure of the IAQ. 
We do this by first constructing a D0L-language for a 1D spin chain and then
use that to generate correspondences on a MCFG 
as in \S \ref{MCFGsec} and \S \ref{spinsMCFGsec}.

\subsubsection{D0L-system and $1$-dimensional quasilattice}
As we recalled in Definition~\ref{D0Lsys} in \S \ref{D0Lsec} of this paper a D0L-system
is a triple $\cD=(\Sigma, \phi, \Omega)$ with $\Sigma$ a finite alphabet,
$\phi: \Sigma^* \to \Sigma^*$ a monoid homomorphism and $\Omega \in \Sigma^*$ 
an initial word or ``axiom".
The D0L-language generated by the system is
$\cL_\cD=\{ \phi^n(\Omega) \,|\, n\in \N \}$. 

\begin{figure}
    \begin{center}
    \includegraphics[scale=0.5]{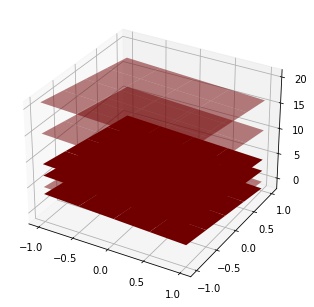}
    \caption{An inflation of a set of planes from the icosahedral Ammann quasilattice (translucent), overlaid on top of the original uninflated quasilattice (opaque). Each set of planes in the quasilattice inflates similarly.}
    \label{fig:inflate}
    \end{center}
\end{figure}

Under the functorial map from formal languages to spin chains, 
the alphabet $\fA$ encodes the spacings of the spin chain and the homomorphism $\Phi$ 
encodes its inflation rules. In the setting of the IAQ, this one-dimensional chain represents,
in the dual space, the spacings between hyperplanes in one system of hyperplanes of the IAQ.
Figure \ref{fig:inflate} shows an inflation of the IAQ in the dual space.

To create the formal language for our three-dimensional quasilattice, 
we first define a D0L-system for a one-dimensional quasilattice that 
describes the spacings, labelled by letters $\bL$ and $\bS$, between
a given family of planes in the IAQ.

\begin{lem}\label{1DIAQ}
Consider the Fibonacci $D0L$ system $\cD=(\Sigma, \phi, \Omega)$ with
$\fA=\{ \bS, \bL \}$, $\Omega=\bL$, and $\phi: \Sigma^* \to \Sigma^*$
given by 
\begin{equation}\label{1D0L-IAQ-Fibonacci}
        \phi(\bL) = \bL \, \, \bS \, , \ \ \ \ 
        \phi(\bS) = \bL \, , 
\end{equation}
The Fibonacci language $\cL_\cD=\{ \Phi^n(\bL) \}_{n\in \N}$ is a context-free language $\cL_\cG$
computed by a context-free grammar $\cG=(\fQ,\fA,P,S)$ with nonterminals $\fQ=\{ S, A, B \}$, 
terminals $\fA=\{ \bS, \bL \}$, and production rules
$$ S\to A \, , \ \ \  A \to AB \, , \ \ \ B \to A \, , \ \ \  A \to \bL \, , \ \  \  B \to \bS \, . $$
\end{lem}

\smallskip

The functor $\fS$ of Theorem~\ref{CFspinFunctor} maps 
the context-free language $\cL_\cD$, together with an assignment
of the interaction terms $J_\fA=\{ J_\bS, J_\bL \}$ 
to a one-dimensional quasilattice and a corresponding spin chain $S_\Upsilon$
with $\Upsilon=(\fA, \cL_{\cD}, J_\fA)$. The substitution rule \eqref{1D0L-IAQ-Fibonacci}
is mapped to an endomorphism of $S_\Upsilon$ in the category $\cS$ of
spin chains. 

\smallskip

It is shown in \S B.1 of \cite{SocStein} that the Fibonacci 1D
quasicrystal, determined by the D0L-formal language with
substitution rules $\bL \to \bL\bS$ and $\bS\to \bL$ as above,
is equivalent to the 1D quasilattice with substitution rules
\begin{equation}\label{1D_FL}
        \phi(\bS) = \frac{\bL}{2}\hspace{0.05cm}\frac{\bL}{2}, \ \ \ \ 
        \phi(\bL) = \frac{\bL}{2}\hspace{0.05cm}\bS\hspace{0.05cm}\frac{\bL}{2}. 
\end{equation}
In this form, a word in $\cL_\cD$ of the form $w=\phi(u)$, for some other word $u\in  \cL_\cD$, is the
1-dimensional chain $S_w$ where each spacing $\bS$ in the chain $S_u$ is replaced 
by a pattern of spacings of the form $\frac{\bL}{2}\frac{\bL}{2}$ and each 
$\bL$ spacing by a pattern $\frac{\bL}{2} \bS \frac{\bL}{2}$. Replacing $S_u$ with $S_w$
corresponds to one inflation of the one-dimensional quasilattice.

\smallskip
\subsection{The Ammann planes quasilattice}\label{AmmannSec2}

\begin{figure}
    \begin{center}
    \includegraphics[scale=0.25]{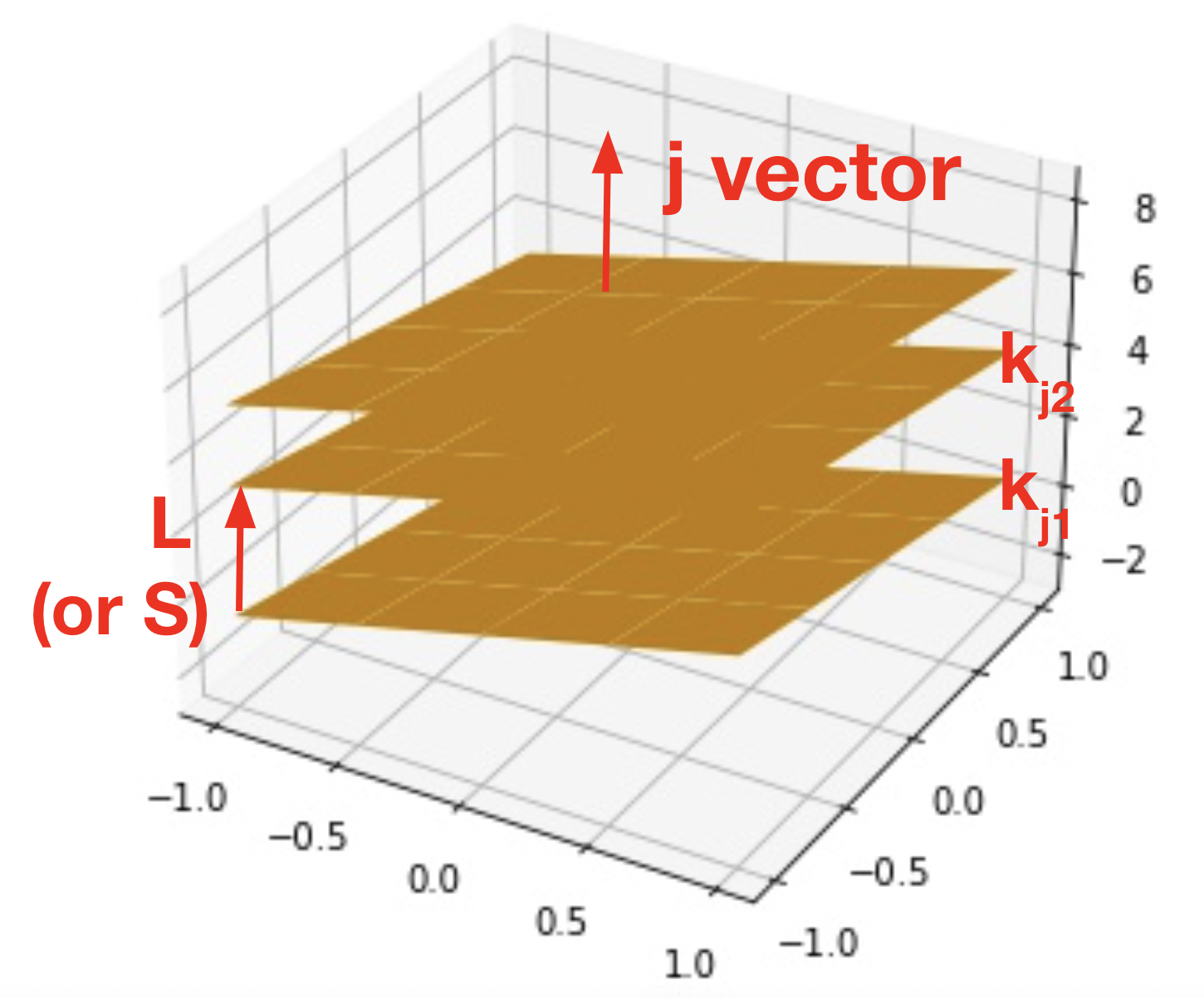}
    \caption{A geometric representation of the $j$-th subsystem of Ammann hyperplanes.  
    The $j$ vector is the icosahedral star vector $\Vec{e}_j$ that a set of planes is orthogonal to. 
    The integers     $k_{j,1}$ and $k_{j,2}$ select two planes in this arrangement and a letter 
    $\bL$ or $\bS$ represents the length of the spacing between them.
    \label{fig:alphabet}}
    \end{center}
\end{figure}

As in \cite{SocStein}, consider the vectors $\{ \Vec{e}_j \}_{j=0}^5$ of the form
\begin{equation}\label{ejvecs}
 \Vec{e}_j =(\frac{2}{\sqrt{5}} \cos(\frac{2\pi j}{5}), \frac{2}{\sqrt{5}} \sin(\frac{2\pi j}{5}), \frac{1}{\sqrt{5}} )\, \text{ for } j=1,\ldots, 4\, , \ \ \ \ \ \ \Vec{e}_5=(0,0,1) \, . 
\end{equation}  
We can then describe the resulting Fibonacci hexagrid as in \cite{SocStein},
as the set of vectors $\Vec{x}$ with the property that the values
$\Vec{x}\cdot \Vec{e}_j$ form a 1D Fibonacci quasi-lattice. Thus, we
can specify the position on this hexagrid in terms of bracketed symbols 
\begin{equation}\label{brasymb}
 \Bigl[\hspace{0.05cm}j\hspace{0.05cm},[\hspace{0.05cm}k_{j_1}\hspace{0.05cm}, \hspace{0.05cm}k_{j_2}\hspace{0.05cm}], \hspace{0.05cm}\Vec{v}_j\hspace{0.05cm}\Bigr] \, , 
\end{equation} 
where each plane in the subsystem of planes normal to the vector $\Vec{e}_j$ is
further labeled by an integer $k_j$, representing its ordinal position along 
the $\Vec{e}_j$ direction. The pair $[ k_{j,1}, k_{j,2} ]$ then identifies a choice of
two planes in the subsystem normal to $\Vec{e}_j$, and $\Vec{v}_j$ represents the translation vector,
see Figure~\ref{fig:alphabet}. After an application of a single substitution rule in the underlying 1D
quasilattice in the $\Vec{e}_j$ direction, 
the $j$ remain unchanged because the planes are orthogonal to the same $\Vec{e}_j$, 
the ordinal positions $k_{j,1}, k_{j,2}$ of the planes have also not changed, while
the translation vector changes, because the IAQ is translated by a 
factor of $\tau$ for each inflation, so that the new translation vector becomes 
$(\tau-1) \Vec{v}_j$ after one inflation and $\tau^n (1-\tau^{-1}) \Vec{v}_j$ after $n$ inflations.

\smallskip
\subsubsection{Multiple context-free grammar and the Ammann quasilattice}

\begin{figure}
    \begin{center}
     \includegraphics[scale=0.4]{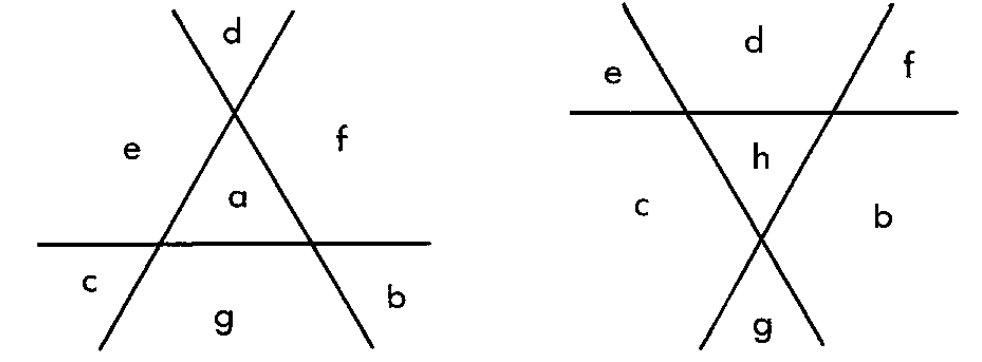} 
    \\
    \includegraphics[scale=0.25]{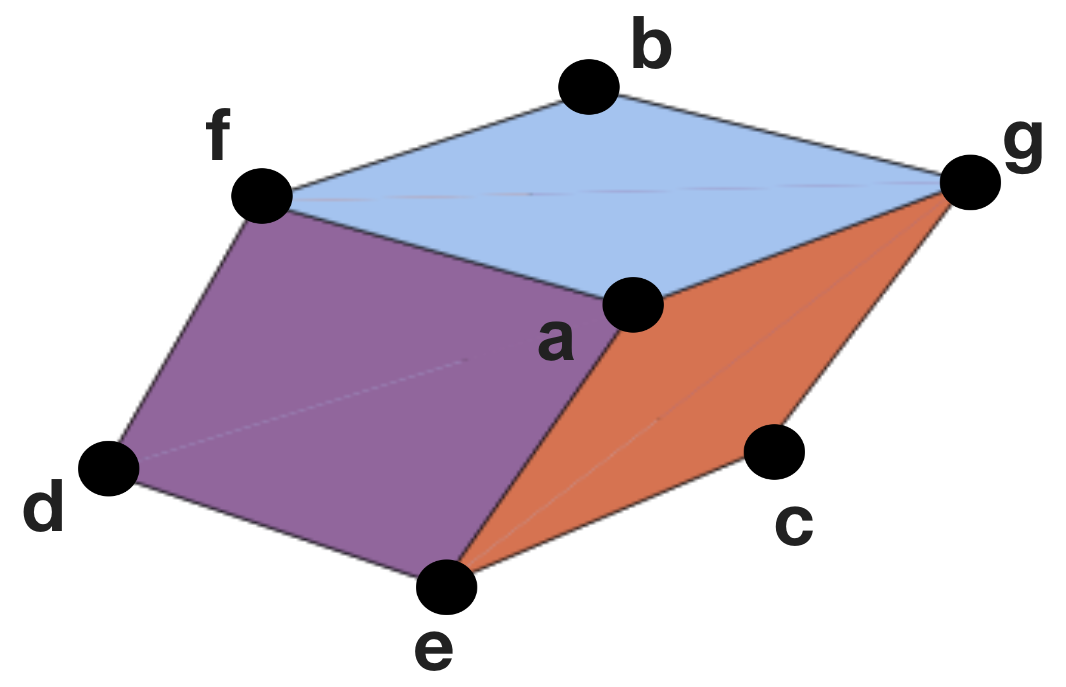} 
    \caption{The eight regions cut out by a parallel pair of $j$-hyperplanes
    cutting near a triple intersection of $\{ j_1, j_2, j_3 \}$ hyperplanes in
    the Ammann quasilattice, and the corresponding eight vertices of a rhombohedron. 
    \label{fig:dualspinsfig}}
    \end{center}
\end{figure}

We are interested here in implementing Korepin's completely integrable model of \cite{Ko}
as a spin system over a quasiperiodic icosahedral structure determined by the 3D Amman
quasilattice described above, seen as an instance of the costruction of spin systems
associated to multiple context free grammars introduced in \S \ref{MCFGsec}. 
To this purpose we will focus on the points of intersections of triples of Ammann 
planes $\{ j_1, j_2, j_3 \}$, for a triple of distinct $j_k\in\{ 0, \ldots, 5 \}$ 
and the eight regions cut out near this intersection point 
by three such hyperplanes and two parallel hyperplanes from another subsystem $j$, see
Figure~\ref{fig:dualspinsfig}. 

\smallskip

We form a multiple context-free grammar that captures this structure of
subsystems of the Ammann quasilattice in the following way.

\begin{prop}\label{IAQmCFG}
Consider the data $\cG=(\fA,\cN,P,S)$ with $\fA=\{ \bL_j ,\bS_j \}_{j=0}^5$ and
$$ \cN=\{ S, M, M_j, N_j, A_j, B_j \,|\, j=0,\ldots, 5 \} $$
with $d(M)=3$, $d(M_j)=d(N_j)=d(A_j)=d(B_j)=1$, $M=(x,y,z)$, 
and with production rules
$$ S \to f(M) \, ,  \ \ \ \  M \to g(M,M_j) \, , \ \ \  M \to h_{j_1,j_2,j_3}( N_0, \ldots , N_5)\, , $$
with $f(x,y,z)=xyz$, $g((x,y,z),u)=(xu,yu,zu)$, and functions 
$$ h_{j_1,j_2,j_3}(x_0,\ldots, x_5)=(x_{j_1},x_{j_2},x_{j_3})\, , $$ 
indexed by triples $(j_1,j_2,j_3)$ of distinct indices $j_k\in \{ 0,\ldots, 5 \}$, and
$$ N_j \to A_j \, , \ \ \ \  N_j \to B_j $$
$$ A_j \to  A_j B_j  \, , \ \ \ \ \   B_j \to A_j   $$
$$ M_j \to \bL_j \, , \ \ \ \  M_j \to \bS_j \, , $$ 
$$ A_j \to \bL_j \, , \ \ \ \ \  B_j \to \bS_j \, . $$
These define a MCFG, where 
\begin{equation}\label{IAQSigmaG}
 \Sigma_\cG=\{ (w_{j_1} \omega_j, w_{j_2} \omega_j, w_{j_3} \omega_j) \, | \, w_{j_k}\in \cL_\cD\, , 
\omega_j\in \{ \bS_j, \bL_j\} \} \, , 
\end{equation}
for any triple $(j_1,j_2,j_3)$ of distinct indices $j_k\in \{ 0,\ldots, 5 \}$ and 
any $j\in \{ 0,\ldots, 5 \}\smallsetminus \{j_1,j_2,j_3 \}$,  with
$\cL_\cD$ the Fibonacci language of Lemma~\ref{1DIAQ}. The words in the
resulting language $\cL_\cG$ parameterize the regions of the 
Ammann quasilattice described above. The substitution $\phi$ of \eqref{1D0L-IAQ-Fibonacci} 
for the Fibonacci D0L-system defines an endomorphism of the MCFG $\cG$ in
the category $\cM\cC\cF\cG_3$. 
\end{prop}

\proof The data above define a MCFG according to Definition~\ref{MCFGdef}. 
The language $\cL_\cG$ consists of the image $f(\Sigma_\cG)$ for $\Sigma_\cG$
as in \eqref{IAQSigmaG}, since the set $\Sigma_\cG$ is the set of triples 
obtained from repeated application of the other production rules. The triple
$(w_{j_1}, w_{j_2}, w_{j_3})$ identifies a triple intersection in the $j_1,j_2,j_3$-subsystem
of Ammann planes, and the additional datum $\omega_j$ specifies the only remaining
information that is needed to identify the eight regions cut out by two parallel hyperplanes
of the $j$-subsystem near this triple intersection, namely whether the spacing
between them is $\bS_j$ or $\bL_j$. The substitution $\phi$ of \eqref{1D0L-IAQ-Fibonacci} 
extends to the MCFG $\cG$ as the substitution $\phi(\bL_j)=\bL_j \bS_j$ and $\phi(\bS_j)=\bL_j$. 
As such it defines a diagonal morphism $X=(X_{ab})_{a,b=1}^3$ acting on triples as
$$ \phi: (w_{j_1} , w_{j_2} , w_{j_3} )\mapsto (\phi(w_{j_1}) , \phi(w_{j_2}) , \phi(w_{j_3}) )\, . $$
If the pair of $j$-hyperplanes cutting out the eight regions around the $j_1,j_2,j_3$-triple
intersection has spacing $\bL$ then after applying the substitution $\phi$ it will have
spacing either $\bL$ or $\bS$, depending on the position of the triple intersection in
the spacing between these two hyperplanes, while if the spacing is $\bS$, it becomes
a $\bL$ spacing after substitution. We refer to the new letter as $\omega_{j,\phi}$ so that
the resulting action of $\phi$ on the triples in $\Sigma_\cG$ is of the form
$$ \phi: (w_{j_1} \omega_j, w_{j_2} \omega_j, w_{j_3} \omega_j)  \mapsto 
(\phi(w_{j_1}) \omega_{j,\phi},  \phi(w_{j_2}) \omega_{j,\phi}, \phi(w_{j_3}) \omega_{j,\phi}) \, . $$
This also defines a morphism in $\cM\cC\cF\cG_3$, given by the graph of $\phi$, 
$$ X_\phi =\{ ((w_{j_1} \omega_j, w_{j_2} \omega_j, w_{j_3} \omega_j), (\phi(w_{j_1}) \omega_{j,\phi},  \phi(w_{j_2}) \omega_{j,\phi}, \phi(w_{j_3}) \omega_{j,\phi}) \}\, , $$
which is no longer diagonal, since $\omega_{j,\phi}$ depends also on 
$(w_{j_1} , w_{j_2} , w_{j_3} )$ not only on $\omega_j$. 
\endproof

\smallskip

\begin{cor}\label{IAQbrack}
When parameterizing the positions in the Ammann quasilattice in terms of the Fibonacci hexagrid 
and the bracketed symbols of \eqref{brasymb}, the substitution rules of
Proposition~\ref{IAQmCFG} can be equivalently described as a mapping $\Phi$ of the form
\begin{equation}\label{3D_FL}
    \Phi(X)\Bigl[\hspace{0.05cm}j\hspace{0.05cm},[\hspace{0.05cm}k_{j_1}\hspace{0.05cm}, \hspace{0.05cm}k_{j_2}\hspace{0.05cm}], \hspace{0.05cm}\Vec{v}_j\hspace{0.05cm}\Bigr] = \phi(X)\Bigl[\hspace{0.05cm}j\hspace{0.05cm}, [ \hspace{0.05cm}k_{j_1}\hspace{0.05cm}, \hspace{0.05cm}k_{j_2}\hspace{0.05cm}], \hspace{0.05cm}(\tau-1)\Vec{v}_j\hspace{0.05cm}\Bigr], 
    \end{equation}
    for $X=\bL$ or $\bS$. For multiple inflations $n$, we have (for $X=\bL$ or $\bS$)
\begin{equation} \label{3D_FL_mult}
    \Phi^{n}(X)\Bigl[\hspace{0.05cm}j\hspace{0.05cm},[\hspace{0.05cm}k_{j_1}\hspace{0.05cm},\hspace{0.05cm}k_{j_2}\hspace{0.05cm}],\hspace{0.05cm}\Vec{v}_j\hspace{0.05cm}\Bigr] = \phi^{n}(X)\Bigl[\hspace{0.05cm}j\hspace{0.05cm}, [\hspace{0.05cm}k_{j_1}\hspace{0.05cm},\hspace{0.05cm}k_{j_2}\hspace{0.05cm}], \hspace{0.05cm}\tau^{n}(1-\frac{1}{\tau})\Vec{v}_j\hspace{0.05cm}\Bigr] \, . 
\end{equation}
\end{cor}

\medskip
\subsection{Exactly solvable spin model on the icosahedral quasicrystal}\label{IcoModelSec}

We show here that the spin system associated to the MCFG of Proposition~\ref{IAQmCFG}
under the functorial mapping of Proposition~\ref{MCFGfunctor} is exactly solvable and
satisfies the Zamolodchikov tetrahedron equation. We also compute the bulk free energy
and show the criticality of the model. The argument we present follows closely the
analogous arguments in  \cite{Bax}, \cite{Bax3}, \cite{Ko}.  After reviewing this case, we 
will discuss how to partially generalize this approach to the spin systems associated to
other more general MCFGs, in \S \ref{SimplexSec}.

\subsubsection{Model on the icosahedral quasicrystal as spin system from a MCFG}

The spin model we consider here is the image under the  functor $\fS$ of 
Proposition~\ref{MCFGfunctor} of the MCFG of Proposition~\ref{IAQmCFG}
together with the datum $J$ as in Proposition~\ref{MCFGfunctor} that
specifies the interaction terms of the Boltzmann weight \eqref{BoltzWm}. 
The following Lemma~\ref{BWeightIAQ} specializes Proposition~\ref{IAQmCFG}
to the specific case of interest here. 

\begin{figure}
    \centering
    \includegraphics[scale=0.4]{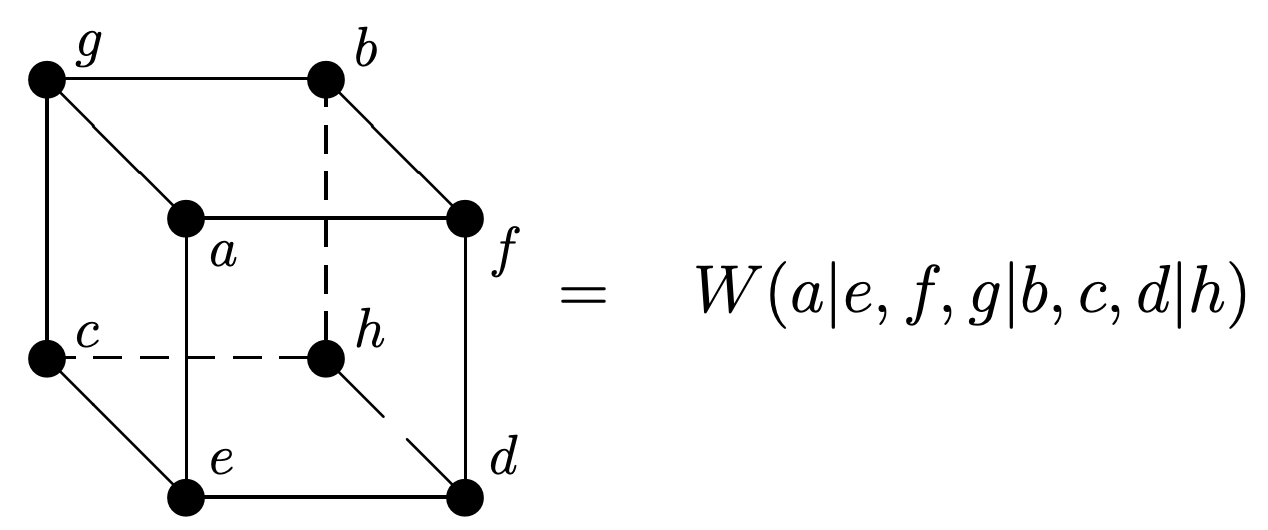}
    \caption{Boltzmann weights for spins at the vertices of the rhombohedron
    dual to the eight regions cut out by a triple $j_1,j_2,j_3$-intersection
    and a pair of $j$-hyperplanes in the Ammann quasi-lattice. \label{FigBWeights} }
\end{figure}

\smallskip

\begin{lem}\label{BWeightIAQ}
The Boltzmann weights for the spin system $\fS(\cG)$, with $\cG$ the MCFG of Proposition~\ref{IAQmCFG}
are of the form $W(a|e,f,g|b,c,d|h)$ on each prolate rhombohedra with spins $a, b, c, d, e, f, g, h$ on its vertices, as in Figure~\ref{FigBWeights}. 
\end{lem}

\smallskip

\begin{figure}
    \begin{center}
    \includegraphics[scale=0.4]{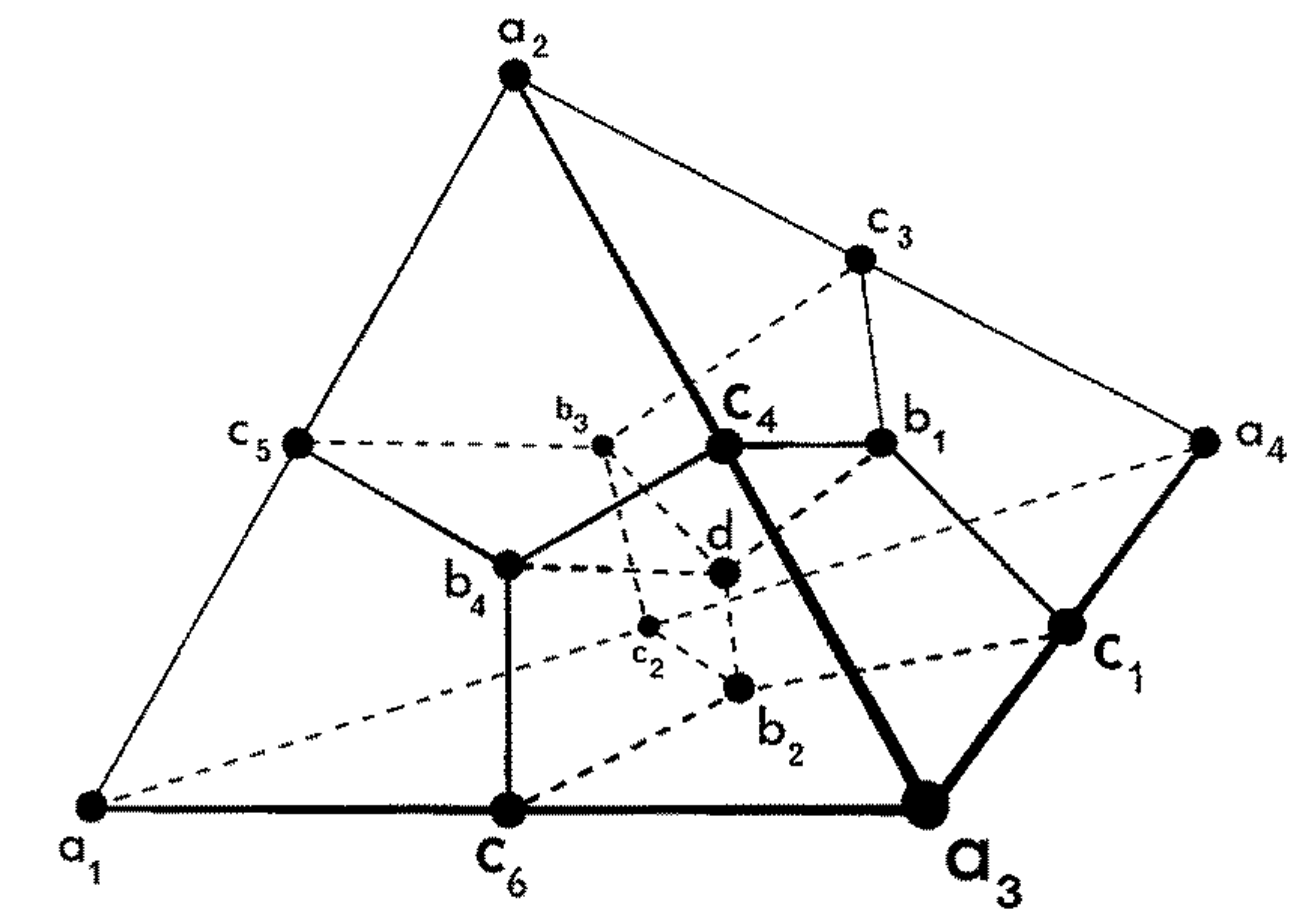}
    \caption{Illustration from \cite{Bax2} of the geometry of the left-hand side of \eqref{transfer_commute}, computing 
    the partition function of four skewed rhombohedra joined together.  As we discuss in \S \ref{SimplexSec}, this is a cubulation of the $3$-simplex. \label{fig:dodecahedron}}
  \end{center} 
\end{figure}

The description above, as given in Proposition~\ref{IAQmCFG}, does not explicitly specify the 
dependence of the Boltzmann weights $W(a|e,f,g|b,c,d|h)$ on parameters describing the spin
interactions, as it allows for different possible models. We assume here that the dependence
on parameters, which specifies the additional datum $J$ in the category $\cM\cC\cF\cG_3\cJ$,
is as assumed in \cite{Bax2}. 
Namely, as solutions of the  Zamolodchikov equations 
\begin{equation} \label{transfer_commute}
    \begin{gathered}
        \sum\limits_d W(a_4|c_2c_1c_3|b_1b_3b_2|d) W'(c_1|b_2a_3b_1|c_4dc_6|b_4) \\ W^{\prime\prime}(b_1|dc_4c_3|a_2b_3b_4|c_5) W^{\prime\prime\prime}(d|b_2b_4b_3|c_5c_2c_6|a_1) \\= \sum\limits_b W^{\prime\prime\prime}(b_1|c_1c_4c_3|a_2a_4a_3|d) W^{\prime\prime}(c_1|b_2a_3a_4|dc_2c_6|a_1)\\ W'(a_4|c_2dc_3|a_2b_3a_1|c_5)W(d|a_1a_3a_2|c_4c_5c_6|b_4)
    \end{gathered}
\end{equation} 
for all values of the fourteen ``external" spins $a_1,...,a_4,b_1,...,b_4,c_1,...c_6$, with
$d$ the ``internal" spin. The equation \eqref{transfer_commute} describes the condition
that the transfer matrices $T(W)$ and $T(W')$ for two given weights $W,W'$ commute,
as the existence of two other weights $W''$ and $W'''$ so that equation \eqref{transfer_commute} holds.
The Zamolodchikov equation \eqref{transfer_commute}
corresponds to the partition function of four skewed rhombohedra joined together with a 
common interior spin $d$ to form a rhombic dodecahedron, as in Figure~\ref{fig:dodecahedron},
from \cite{Bax2}. The right-hand side of \eqref{transfer_commute} is obtained by 
connecting $d$ to $a_1,...,a_4$ rather than $b_1,...b_4$, so that each term $W,W',W'',W'''$
on the left hand side has a corresponding term on the right-hand-side where $a_i \leftrightarrow b_i$
and $c_1\leftrightarrow c_5$, $c_2 \leftrightarrow c_4$ and $c_3 \leftrightarrow c_6$.
Note that these are reflections about the center of the barycentric subdivision of the $3$-simplex of
Figure~\ref{fig:dodecahedron}. We will return to give a more general geometric interpretation of
\eqref{transfer_commute} in Theorem~\ref{CubeThm} where we extend it for the
Boltzmann weights of the form \eqref{BoltzWm}.

The Boltzmann weights are also assumed in \cite{Bax2} to satisfy the symmetries
\begin{equation} \label{negate}
    W(-a|-e,-f,-g|-b,-c,-d|-h) = W(a|e,f,g|b,c,d|h)\, , 
\end{equation}
$$ W(-a| e,f,g |-b,-c,-d| h) =W(a|-e,-f,-g|b,c,d|-h) =W(a|e,f,g|b,c,d|h) \, . $$

\medskip

In order to complete our proof that the spin system on the icosahedral quasicrystal is indeed 
an example of the class of spin systems from MCFGs obtained as in Proposition~\ref{MCFGfunctor},
we need to specify the datum $J$ in $\cM\cC\cF\cG\cJ$, namely the parameters $\underline{J}_{\fA^3}$
of the model. 

\smallskip

\begin{prop}\label{Jparams}
The datum $J$ for the construction of the spin model on the icosahedral quasicrystal as in 
Proposition~\ref{MCFGfunctor}, with the MCFG of Proposition~\ref{IAQmCFG}, is the function
that assigns to a triple in $\fA^3$, or equivalently to a type of rhombohedral region 
determined by a triple intersection in the Ammann quasilattice, the set of spectral parameters
$\theta_{1}, \theta_{2}, \theta_{3}$ obtained from the unit vectors $\Vec{n}_{j_1}$,  $\Vec{n}_{j_2}$,
$\Vec{n}_{j_3}$ associated with the corresponding hexagrid vectors $\Vec{e}_{j_1}$, $\Vec{e}_{j_2}$,
$\Vec{e}_{j_3}$,
\begin{equation}\label{njangles}
    \begin{gathered}
        \Vec{n}_j \cdot \Vec{n}_k=\cos\theta_{1}, \ \ \ \  \Vec{n}_i \cdot \Vec{n}_k=\cos\theta_{2},\ \ \ \  
        \Vec{n}_i \cdot \Vec{n}_j =\cos\theta_{3} \, .
    \end{gathered}
\end{equation}
The Boltzmann weights are a function of these parameters,
\begin{equation}\label{BoltzTheta}
    W_{\underline{J}}(\sigma) = W(a|e,f,g|b,c,d|h; \theta_{1}, \theta_{2}, \theta_{3}) \, ,
\end{equation}    
with the notation of \S \ref{spinsMCFGsec} on the left-hand-side and the notation of  \cite{Bax2} on the
right-hand-side. 
\end{prop}

\proof The spectral parameters are introduced in \cite{Bax2}. The angles  
$\theta_{1}, \theta_{2}, \theta_{3}$ can be seen as the angles of a spherical triangle. 
To show the explicit dependence of the Boltzmann weights on these parameters,
one can introduce, as in \cite{Bax2}, the spherical excesses
\begin{equation}\label{excesses}
    2\alpha_0 = \theta_{1} + \theta_{2} + \theta_{3} - \pi \, , \ \ \  \alpha_\ell = \theta_{l} - \alpha_0\, ,
\end{equation}
where $\ell = 1,2,3$, and define
\begin{equation}\label{ctseq}
    \begin{gathered}
        t_\gamma= \tan(\alpha_\gamma/2)^{1/2}, \ \ \  s_\gamma = \sin(\alpha_\gamma/2)^{1/2}, \\
        c_\gamma=\cos(\alpha_\gamma/2)^{1/2}, \ \ \  \gamma\in\{0,1,2,3\}
    \end{gathered}
\end{equation}
and take, following the same notation used in \cite{Bax2}, 
\begin{equation}
\begin{gathered}
    P_0=1, \ \ \  Q_0=t_0t_1t_2t_3, \ \ \  R_0=\frac{s_0}{c_1c_2c_3}, \\
    P_\ell =t_mt_n, \ \ \ Q_\ell =t_0t_\ell , \ \ \  R_\ell=\frac{s_\ell}{c_0c_mc_n}  \, ,
\end{gathered}
\end{equation}
for $(\ell,m,n)$ a permutation of $(1,2,3)$, as in (3.12) and (3.13) of \cite{Bax2}. 
It is then shown in \cite{Bax2} that the Boltzmann weights \eqref{BoltzTheta}
are completely determined by the values of three spin products $abeh, acfh, adgh$, 
with values given in Table~\ref{Tab1} (as in \cite{Bax2}).
\endproof

\begin{table}\label{Tab1}
\begin{center}
\begin{tabular}{||c c c c||} 
 \hline
 $abeh$ & $acfh$ & $adgh$ & $W(a|e,f,g|b,c,d|h;\theta_{1},\theta_{2},\theta_{3})$ \\ [0.5ex] 
 \hline\hline
 + & + & + & $P_0 - abcdQ_0$ \\ 
 \hline
 - & + & + & $R_1$ \\
 \hline
 + & - & + & $R_2$ \\
 \hline
 + & + & - & $R_3$ \\
 \hline
 + & - & - & $abP_1 + cdQ_1$ \\
 \hline
 - & + & - & $acP_2 + bdQ_2$ \\
 \hline
 - & - & + & $adP_3 + bcQ_3$ \\
 \hline
 - & - & - & $R_0$ \\ 
 \hline
\end{tabular}
\bigskip
\caption{The values from \cite{Bax2} 
of the Boltzmann weights of the spin model, given the signs of the spin products $abeh, acfh, adgh$.}
\end{center}
\end{table}

\medskip

\begin{figure}
    \begin{center}
        \includegraphics[scale=0.3]{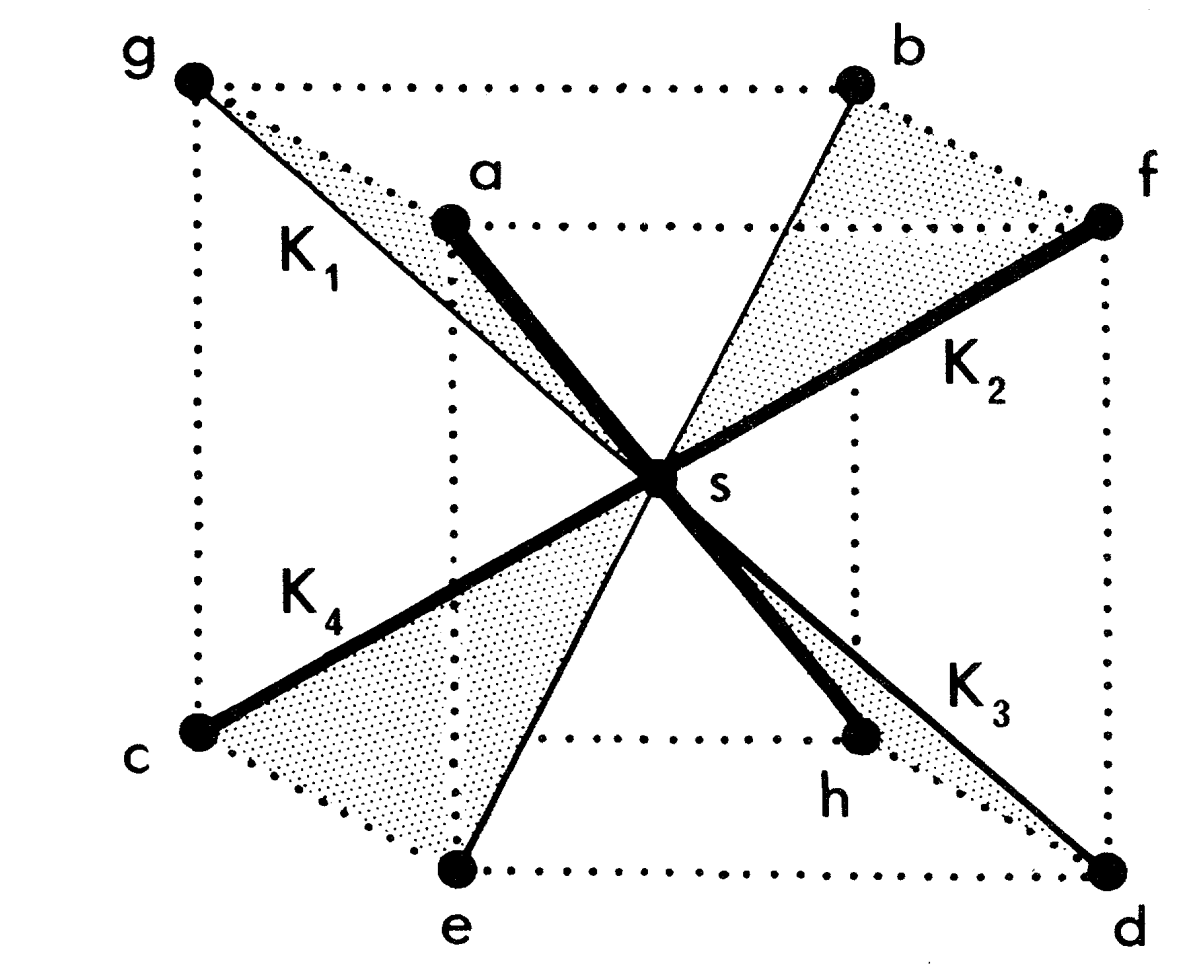}
    \caption{The four areas (represented by shaded triangles)
    over which three-spin interactions $s \sigma_j\sigma_t$ occur, from \cite{Bax3}.
    \label{fig:spinInteractions}}
    \end{center}
\end{figure}

We can also use the properties of the spin model on the icosahedral quasicrystal to
clarify the role of the parameters in the datum $J$ in the mapping of Proposition~\ref{MCFGfunctor}.

\begin{prop}\label{JsProp}
In Proposition~\ref{MCFGfunctor} different possible assignments of the $J$ datum, for the
same underlying MCFG $\cG$ can map to the same spin model under different but equivalent
descriptions. For example, in the case of the spin model on the icosahedral quasicrystal,
one can either assign $J$ to consist of the parameters $\theta_1,\theta_2,\theta_3$ as in
Proposition~\ref{Jparams} above, or one can assign $J$ to be, for each type of rhombohedral region
associated to a triple intersection in the Ammann quasilattice (each element in $\fA^3$), the
set of parameters $K_1, K_2,K_3,K_4$ in the description of the Boltzmann weights in terms
of transfer matrix and body-center-cube lattice of  \cite{Bax3}, depicted in Figure~\ref{fig:spinInteractions}.
These are the coefficients of triple interaction terms of three-spin $s \sigma_j\sigma_t$, with
a spin $s$ placed at the center of every rhombohedron as in the Figure, with the triple
interactions corresponding to the four shaded regions.
The $K_1, K_2,K_3,K_4$  are related to the angle parameters by the following transformations (from \cite{Bax3}):
\begin{equation} \label{K}
    2K_1 = -x'-iy',    \ \ \ \ \     2K_2 = x-iy,     \ \ \ \ \ 
    2K_3 = -x'+iy,     \ \ \ \ \     2K_4 = x+iy
\end{equation}
where 
\begin{equation} \label{x}
      \tanh(x) = t_0t_3,  \ \ \ \ \  \tan(y) = t_2/t_1, \ \ \ \ 
        \tanh(x') = t_1t_2, \ \ \ \ \tan(y') = t_0/t_3 \, , 
\end{equation}
with the $t_i$ as in \eqref{ctseq}.
\end{prop}

\proof As in \cite{Bax3}, for a cube (or a rhombohedron, as the cube-cells in the 
cubulation of the $3$-simplex) consider the regions
depicted in Figure \ref{fig:spinInteractions}, with 
a spin located at the center of the rhombohedron. Then, looking at the shaded 
triangles in Figure~\ref{fig:spinInteractions}, define $K_1,\ldots, K_4$ as stated,
as the interaction coefficients for each three-spin interaction $s_i\sigma_j\sigma_t$.
The parameters $K_1,\ldots ,K_4$ are not independent 
of each other, but because the spins in each layer only interact with the $s$ spins above and below them—and the spins above do not interact with the spins below—the transfer matrix $T$ can be factored,
as shown in \cite{Bax3}, in the form $T = \xi^n X(K_3,K_4)Y(K_1,K_2)$,
where the top spins are accounted for through $Y$ and the bottom spins are accounted for 
through $X$. The relation to the previous expression of the Boltzmann weights is through the
partition function and its dependence on the $K_1,\ldots, K_4$ parametrs. It is shown in \cite{Bax3}
that the partition function of the spin system can be written as
\begin{equation}\label{Z_BFEP}
    Z = \xi^N \Tr(R(v_3)S(v_1))^{k_j} \, ,
\end{equation}
for 
$$ \xi = \frac{1}{2}\gamma(\frac{1}{2}\sin\theta_{3})^{1/2} $$
and for diagonal matrices $R(u),S(v)$ depending on $u,v \in \C$, which satisfy
$X(u) = PR(u)Q^{-1}$ and  $Y(v) = QS(v)P^{-1}$, 
for nonsingular matrices $P, Q$, where $X$ and $Y$ are as above, where the
dependence on the parameters $K_1,K_2,K_3,K_4$ is reduced to $X$ as a function 
$X(v_3)$ and $Y$ as a function $Y(v_1)$, for $T_i = \tan(\theta_i/2)^{1/2}$, $z = \exp(ia_3/2)$,
$v_i = \tanh(2K_i), i=1,...,4$, and 
$v_1 = -zT_1T_2, v_2 = -izT_2/T_1$, $v_3 = -z^{-1}T_1T_2, v_4 = iz^{-1}T_2/T_1$, 
 $v_1v_4 + v_2v_3 = 0$, 
        so that if $T_1$ is fixed, then $v_4 = -iv_3/T_1^2$, and $X$ depends on $\theta_2$ and $\theta_3$ 
only via $v_3$, and similarly $Y$ depends on them only via $v_1$.
The exponent $k_j$ in \eqref{Z_BFEP} denotes the number of planes in the subsystem
orthogonal $\Vec{e}_j$, or equivalently the length of a word in the $\cL_\cD$ Fibonacci 
language associated to the $j$-subsystem of Ammann planes, as we will discuss
further in \S \ref{freeSec}.
\endproof

\subsubsection{Boltzmann weights and solvability in the icosahedral quasicrystal model}\label{solvSec}

We recall here the descriptions of the Boltzmann weights, as in \cite{Bax2}, 
in terms of Zamolodchikov spins associated to the twelve faces separating the eight
regions described above

\medskip

\begin{figure}[h]
    \begin{center}
    \includegraphics[scale=0.2]{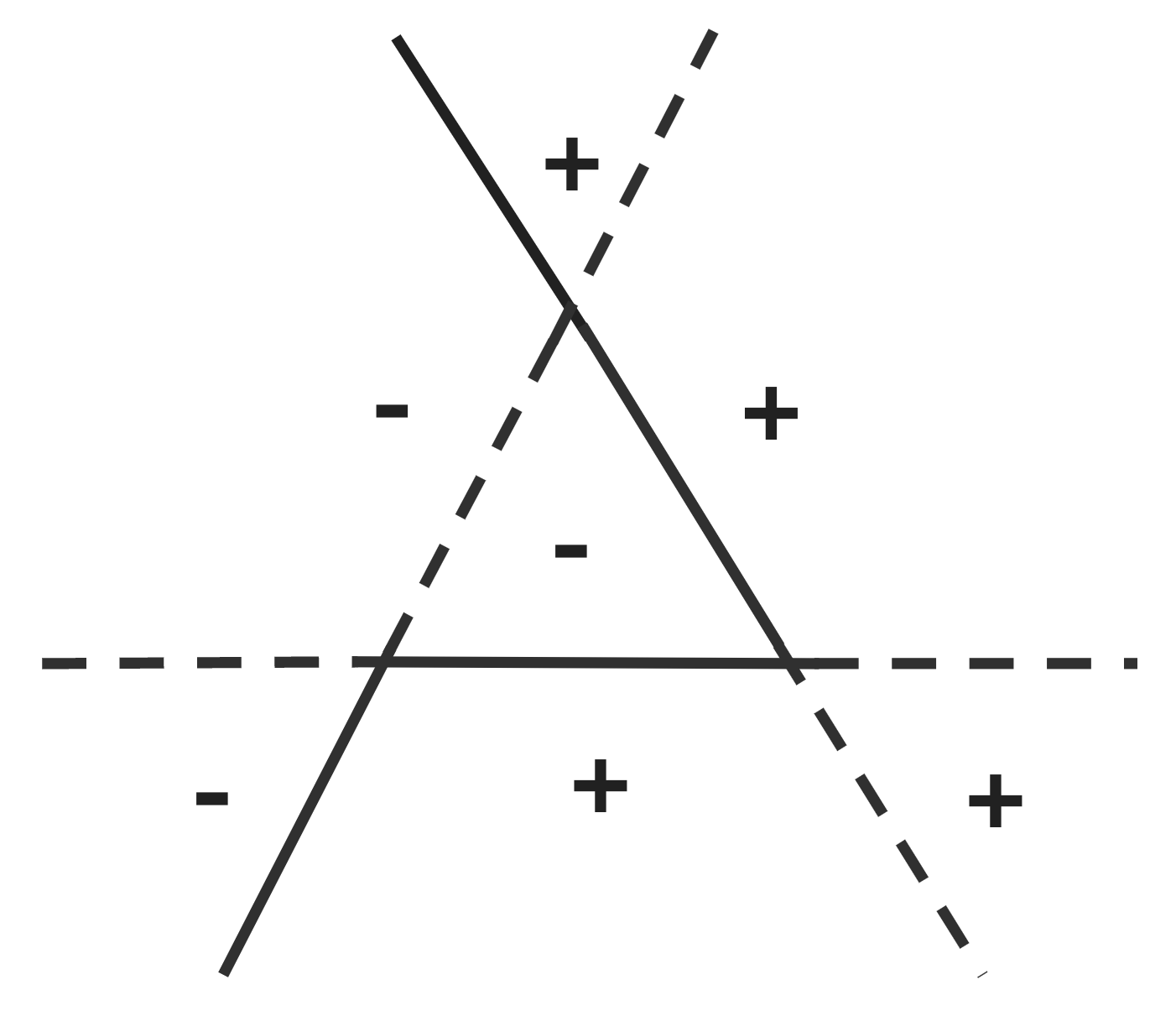}
       \caption{Face colorings, represented by the dashed and straight lines, of the 
    polygons comprising the polyhedra in the IAQ. The faces are dashed lines if the 
    spins on both sides agree and straight lines otherwise, see \cite{Zamo}. 
    \label{fig:faceColorings}}
    \end{center}
\end{figure}

In the dual space, spins surrounding a threefold singularity are separated by faces of the 
icosahedral Ammann quasilattice. 
Recall that a face of the quasilattice is defined as the polygons comprising each polyhedra in the dual space. Let us color each face white if the spins on either side are equal, and black if they are different (Figure \ref{fig:faceColorings}). Then by letting spins $a,...,h$ take on all possible values, one
 obtains the allowed colorings in Zamolodchikov's \cite{Zamo}. From \eqref{negate}, $W$ only depends on the colorings of the faces. Therefore we can replace the function $W$ of eight spins with a function $S$ of the colors of the twelve faces. If the color white is $+$ and the color black is $-$, not to be confused with the $+$ and $-$ denoting positive and negative spin values in Figure \ref{fig:faceColorings}, then the color on the face between two spins $c$ and $g$ is the product $cg$ of the spins. Thus,
\begin{equation}\label{SspinW}
    S(cg, ae, df, bh, de, af, bg, ch, bf, ag, ce, dh) = W(a|e,f,g|b,c,d|h),
\end{equation}
where $S$ are Zamolodchikov's spins, hence, \eqref{transfer_commute} 
is precisely Zamolodchikov's tetrahedron equations, as in \cite{Zamo}. 
This formulation is used to show that the model is exactly solvable.

\smallskip
\subsection{Spin systems from MCFGs and the cubulation of the $n$-simplex}\label{SimplexSec}

We have reviewed some of the main properties of the completely
integrable model on the icosahedral quasicrystal and its relation to
the Zamolodchikiv tetrahedron equations.
We now consider again the general case of Definition~\ref{kspinsys}. We show that the
tetrahedron equation in the form \eqref{transfer_commute} has a natural analog in this more general
setting, for Boltzmann weights of the form \eqref{BoltzWm}, and is associated to a dual
pair of cubulations of the $n$-simplex. The cubulation of the $n$-simplex is the analog, 
in this higher dimensional setting, of the 3-simplex decomposed into four rhombohedra 
(cubes) as in Figure~\ref{fig:dodecahedron}. 

\smallskip

A cubulation of a topological space 
is a decomposition into a union of cubes that intersect only along faces (of any codimension).
Note that (as in the case of the rhombohedra of Figure~\ref{fig:dodecahedron}) the cubes
of the cubulation are not necessarily isometric to cubes, but are homeomorphic images of
cubes (that is, their shapes can be distorted): they retain the combinatorial structure of cubes,
in terms of number of vertices and faces and their incidence relations, with faces of dimension $k$ 
homeomorphic to $k$-cubes. 

\smallskip

\begin{lem}\label{cubulations}
The simplex $\Delta_n$ has two dual cubulations $\cC(\Delta_n)$ and $\cC(\Delta_n')$,
both consisting of $(n+1)$ cubes of dimension $n$. In  $\cC(\Delta_n)$ each cube is the 
union of all the simplexes of the barycentric subdivision $\cB(\Delta_n)$ that are incident to 
the same vertex of $\Delta_n$ and in $\cC(\Delta_n')$ each cube is the union of all the
simplexes of the barycentric subdivision that are incident to the same vertex of $\cB(\Delta_n)$
located in the middle of one of the $(n-1)$-dimensional faces of $\Delta_n$.
\end{lem}

\proof 
The cubulation $\cC(\Delta_n)$
of the $n$-simplex can be obtained, as in \cite{ShtSht}, by embedding the $n$-simplex
$\Delta_n$ inside the cube $\cI^{n+1}=[0,1]^{n+1}$ as the set
$\Delta_n=\{ (x_0,\ldots, x_n)\in \cI^{n+1} \,|\, x_0+\cdots + x_n=1 \}$. Consider then
the set of $n$-dimensional faces $F_i$ of the cube $\cI^{n+1}$ that lie on the hyperplanes $x_i=1$.
These are half of the faces of the $(n+1)$-cube. The projection of the union $\cup_i F_i$ of these
faces onto the simplex $\Delta_n\subset \cI^{n+1}$ along the rays from the vertex 
$(0,0,\ldots, 0)\in \cI^{n+1}$ determines a cubical decomposition of the simplex $\Delta_n$. 
The cubes of this cubulation $\cC(\Delta_n)$ are equivalently described (see \cite{ShtSht}) as the union
of the simplexes of the barycentric subdivision $\cB(\Delta_n)$ of $\Delta_n$ that are incident to the same
vertex of $\Delta_n$.
This construction of the cubulation $\cC(\Delta_n)$ of the simplex is the same as 
the {\em pair subdivision} used by Sullivan and Rounds, see \cite{Rounds}: in particular, 
given a triangulation of an $n$-manifold
and its dual triangulation, the intersections of any simplex in the triangulation with the
cells of the dual triangulation gives a cubulation of the simplex, which is equivalent
to the one described above. 

\begin{figure}[h]
    \begin{center}
    \includegraphics[scale=0.2]{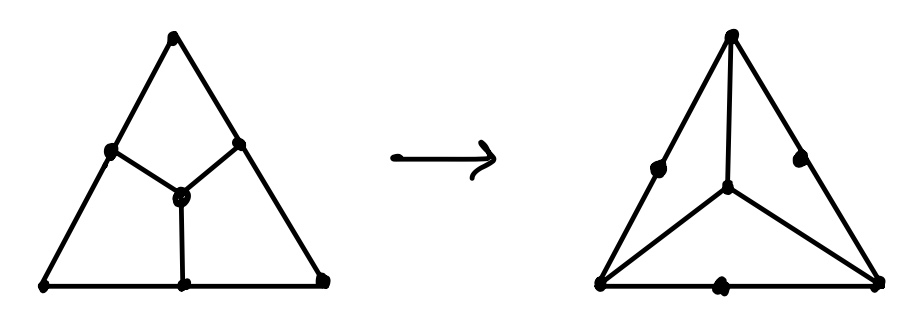} \\
    \includegraphics[scale=0.2]{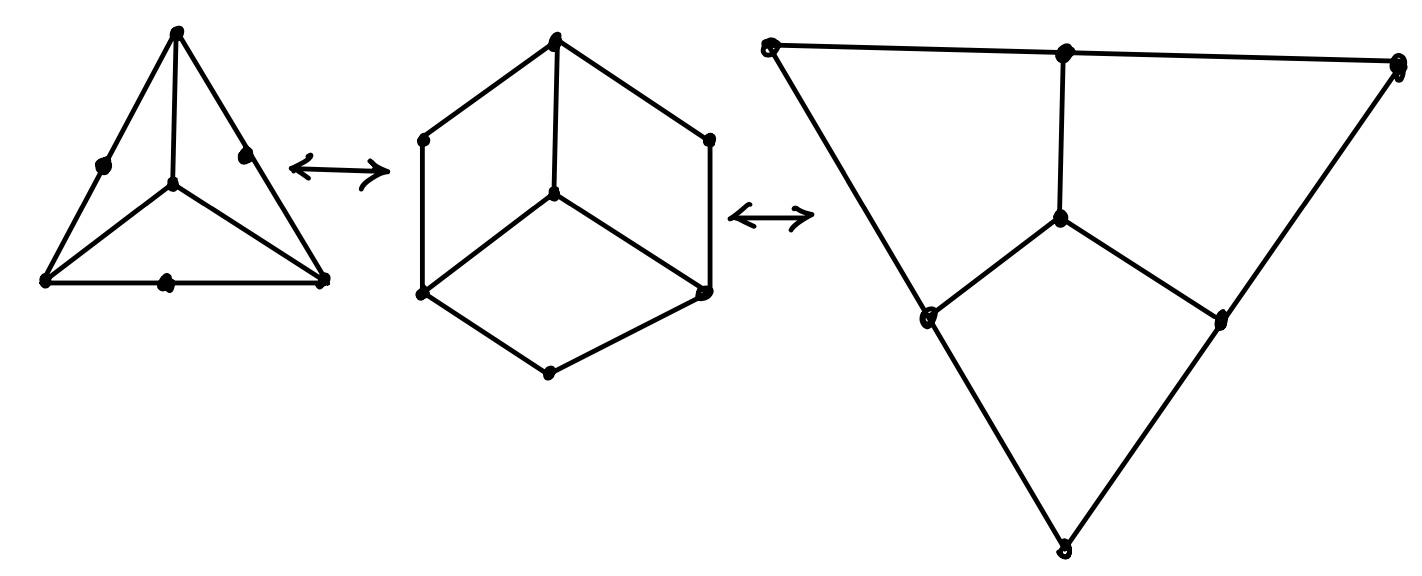}
       \caption{The two cubulations $\cC(\Delta_2)$ and $\cC(\Delta'_2)=\cS(\Delta_2)$. 
    \label{2CubulationFig}}
    \end{center}
\end{figure}

We can then consider a different decomposition of  $\Delta_n$, which we denote by
$\cS(\Delta_n)$, where each cell of the decomposition is the union of all the simplexes of
the barycentric subdivision $\cB(\Delta_n)$ that are incident to a vertex of $\cB(\Delta_n)$
located in the middle of one of the $(n-1)$-dimensional faces of $\Delta_n$. This gives
a decomposition of $\Delta_n$ into $n+1$ cells all of which are incident to the central
vertex of the barycentric subdivision. Moreover, this decomposition can also be identified
with a cubulation of an $n$-simplex $\Delta'_n$, where the new $n$-simplex $\Delta'_n$ 
now has as vertices the $n+1$ vertices in the middle of the faces of the original simplex
$\Delta_n$. The regions constructed above are the cubes of this cubulation, so
$\cS(\Delta_n)=\cC(\Delta'_n)$. We refer to it as the {\em dual cubulation}. 
This is illustrated in Figure~\ref{2CubulationFig} in
the case of a $2$-simplex, and one can derive from Figure~\ref{fig:dodecahedron}
the analogous construction for the $3$-simplex. 
\endproof

\medskip

\begin{prop}\label{indeppaths}
In an $n$-dimensional cube, there are exactly $n$ independent paths
that start at the vertex $v_0$ labelled by $\underline{\epsilon}(0)$ and end at the 
vertex $v_{n+1}$ labelled by $\underline{\epsilon}(n)$ and are otherwise vertex disjoint.
In the cubulation $\cC(\Delta_n)$, let $v$ be the vertex common to all the $n+1$ cubes.
For any cube $C$ in the cubulation label the vertices so that $v_n=v$ (so $v_0$ is 
a vertex of $\Delta_n$). Any of the $n$ independent paths of $C$ from 
$v_0$ to $v_n$ will be incident to all the $n+1$ cubes in the cubulation. 
\end{prop}

\proof
The first statement on the $n$ independent paths 
follows from the $n$-connectedness property of the $n$-cube, \cite{Balinski}
and from the characterization of $n$-connectedness as the property that there are
at least $n$ independent paths between any two vertices, \cite{Gor}.
Since $v_0$ has only $n$ neighboring vertices the number of such paths is exactly equal to $n$.
We denote by $\gamma^{i}$, for $i=1,\ldots, n$ these $n$ paths and by  $v^{(i)}_k$, 
for $k=1,\ldots, n-1$, their intermediate vertices, which belong to the set 
of vertices labelled by $\underline{\epsilon}(k)$. 

For the second statement, consider an $n$-simplex $\Delta_n$ with the cubulation
by $n+1$ $n$-dimensional cubes as described in Lemma~\ref{cubulations}. The vertex $v$
that is shared by all of the $n+1$ cubes is the image of the vertex $(1,1,\ldots,1)\in \cI^{n+1}$
in the description of the cubulation as projection from  $\cI^{n+1}$ in Lemma~\ref{cubulations}.
Take one of these $n$-cubes and identify $v=v_n$, with $v_0$ the opposite
vertex, the only vertex of the cube that is also a vertex of the simplex $\Delta_n$. 
Let $\gamma^i$ be
the $n$ independent paths described above, in this $n$-cube. We show
inductively that, for each of these paths, all the $n+1$ cubes in the cubulation will be incident
to at least one of the vertices of the path. So the path touches all the cubes in
the decomposition. This can be seen by direct inspection of Figure~\ref{fig:dodecahedron}
in the case of the cubulation of the $3$-simplex, and even more easily in the case 
of the $2$-simplex. Assume that this is also the case for the $(n-1)$-dimensional simplex
$\Delta_{n-1}$. The statement is equivalent to the corresponding $n-1$ paths in the union
$\cup_i F^{(n)}_i$ of the faces $F^{(n)}_i\subset \{ \underline{x}\in \R^n_{\geq 0}\,|\, x_i=1 \}$, for $i=0,\ldots, n-1$, 
of the cube $\cI^n$ touching every face $F^{(n)}_i$. Now consider the simplex $\Delta_n\subset \cI^{n+1}$.
It has the same property provided that the $n$ independent paths in any one of the faces $F_i^{(n+1)}$
touches all the other faces. We consider $\cI^n \subset \cI^{n+1}$ with the embedding
$(x_0,\ldots, x_{n-1})\mapsto (x_0,\ldots, x_{n-1},0)$
The faces $F_i^{(n+1)}$ are given by $F_i^{(n+1)}=F_i^{(n)}\times [0,1]$ for $i=0,\ldots, n-1$ and
$F_n^{(n+1)}=\cI^{n+1}\cap \{ x_n=1 \}$. the $n$ independent path in the $F_i^{(n+1)}$ faces,
for $i=0,\ldots, n-1$ can be obtained by taking the first $n$ steps of the $n-1$ independent paths of 
the corresponding $F_i^{(n)}$ faces, followed by an edge along the $[0,1]$ interval in
$F_i^{(n)}\times [0,1]$, followed by the last edge of the paths in $F_i^{(n)}$ but embedded in $\cI^{n+1}$
with last coordinate $x_n=1$, so that it is along the intersection of $F_i^{(n+1)}$ with the $F_n^{(n+1)}$ face.
By symmetry it then follows that the independent paths in $F_n^{(n+1)}$ also have the same property,
as we can take any other embedding $\cI^n \subset \cI^{n+1}$, with respect to any
of the other coordinates, which will identify the face $F_n^{(n+1)}$ with one obtained
as the product of an $F_i^{(n)}$ and an interval $[0,1]$, and repeat the same argument. 
\endproof

\medskip

We can then use the dual cubulations and their sets of independent paths to construct 
generalization of the equation \eqref{transfer_commute}. 

\begin{thm}\label{CubeThm}
The equation \eqref{transfer_commute} extends to Boltzmann weights of the form \eqref{BoltzWm}
as an identity
\begin{equation}\label{IdWn}
\begin{gathered}
\sum_{v} \prod_{k=0}^n W_{(k)}(\sigma_{v_{k,0}}| (\sigma_{v^i_{k,1}})_{i=1}^n | \cdots | (\sigma_{v^i_{k,n-1}})_{i=1}^n | \sigma_{v_{k,n}})  = \\ \sum_{v} \prod_{k=0}^n W_{(k)}(\sigma_{\hat v_{k,0}}| (\sigma_{\hat v^i_{k,1}})_{i=1}^n | \cdots | (\sigma_{\hat v^i_{k,n-1}})_{i=1}^n | \sigma_{\hat v_{k,n}})\, ,
\end{gathered}
\end{equation}
where the $v^i_{k,r}$ and $\hat v^i_{k, r}$ are the vertices of the independent paths $\gamma^i$ of Proposition~\ref{indeppaths} in the two dual cubulations $\cC(\Delta_n)$ and $\cC(\Delta_n')$ of Lemma~\ref{cubulations}, where in each term $$W_{(k)}(\sigma_{\hat v_{k,0}}| (\sigma_{\hat v^i_{k,1}})_{i=1}^n | \cdots | (\sigma_{\hat v^i_{k,n-1}})_{i=1}^n | \sigma_{\hat v_{k,n}})$$ the vertices $\hat v^i_{k,r}$ correspond to the
opposite vertices, with respect to the center of the barycentric subdivision, of the $v^i_{k,r}$ in
the corresponding 
term $$W_{(k)}(\sigma_{v_{k,0}}| (\sigma_{v^i_{k,1}})_{i=1}^n | \cdots | (\sigma_{v^i_{k,n-1}})_{i=1}^n | \sigma_{v_{k,n}}) \, . $$ 
\end{thm}

\proof 
Choose one of $n+1$ cubes $C_0$ in the cubulation of the $n$-simplex $\Delta_n$ and
consider the $n$ independent paths, which we denote by $\gamma_0^i$ in this cube, with 
ending vertex $v_{0,n}=v$
the vertex common to all $n+1$ cubes and with starting vertex $v_{0,0}$ the vertex of the $n$-simplex 
that is in $\Delta_n\cap C_0$, as above, and with $v^i_{0,k}$ the intermediate vertices. For each of the
intermediate vertices, identify a corresponding other cube $C_k$ of the cubulation
such that  $v^i_{0,k}$ is a vertex of $C_k$ and $C_k \neq C_j$ for $k\neq j$, so that
all the cubes are accounted for. This is possible because, as shown in Proposition~\ref{indeppaths},
each path $\gamma^i$ in $C_0$ touches all the cubes. For each $C_k$, for
$k=1,\ldots, n$ similarly consider the $n$ independent paths, which we denote
by $\gamma_k^i$, with starting vertex $v_{k,0}=v^i_{0,k}$ and ending at the opposite
vertex $v_{k,n}$ in the cube $C_k$, and with intermediate vertices $v^i_{k,r}$, $r=1,\ldots, n-1$. 
The left-hand-side of \eqref{transfer_commute} can then be phrased, for the Boltzmann weights 
$W^{(k)}(\sigma_{\underline{\epsilon}})$ of \eqref{BoltzWm}, as 
\begin{equation}\label{sumWn}
\sum_{v} \prod_{k=0}^n W_{(k)}(\sigma_{v_{k,0}}| (\sigma_{v^i_{k,1}})_{i=1}^n | \cdots | (\sigma_{v^i_{k,n-1}})_{i=1}^n | \sigma_{v_{k,n}}) 
\end{equation}
where the vertex $v$ is where the $(n+1)$-cubes of the cubulation of
of the $n$-simplex are joined, and the sum is over the total number of simplexes considered. 
To construct the analog of the right-hand-side of \eqref{transfer_commute}, we consider the
cubes of the dual cubulation $\cC(\Delta'_n)$. The right-hand-side of
\eqref{transfer_commute} is the same as the left-hand-side, but applied to this dual cubulation with
the vertices reflected about the center of the barycentric decomposition in the corresponding terms.
Thus, we can generalize the identity \eqref{transfer_commute}  as the identity \eqref{IdWn},
where the $\hat v^i_{k,r}$ are the vertices on the independent paths in the dual cubulation,
related by reflection about the center of the barycentric decomposition. 
\endproof

\medskip

There are generalizations (both classical and quantum) of the Zamolodchikov tetrahedron equation to
simplex equations in higher dimension, see for instance \cite{Bard}, \cite{DiMuHo}, \cite{FreMoo}, \cite{KoRi}.
We can regard \eqref{IdWn} as a form of $n$-simplex equation generalizing the
tetrahedron equation written in the form \eqref{transfer_commute}.

\medskip
\subsubsection{Bulk free energy}\label{freeSec}

We return to the case of the icosahedral quasicrystal with the Boltzmann weights as in
Lemma~\ref{BWeightIAQ} and Figure~\ref{FigBWeights}, and with the 
description of the Boltzmann weights as in \cite{Bax3}, in 
terms of the transfer matrix and body-centered-cube (BCC) lattice that we
already recalled in Proposition~\ref{JsProp}. We discuss here the bulk free energy
calculation and we discuss the analogous question for more general spin systems
associated to MCFGs.

\smallskip

The formulation of  \cite{Bax3} in terms of the transfer matrix $T=T(W)$, is based on the relation
$Z = \Tr(T^{k_j})$, as in \cite{JaeMai}, with $k_j$  the number of planes in the subsystem
orthogonal to $\Vec{e}_j$, or equivalently the length of a word in the $\cL_\cD$ Fibonacci 
language associated to the $j$-subsystem of Ammann planes. 
 From the description \eqref{3D_FL} of the IAQ, with $k_j$ as above, 
if each layer has $n$ threefold singularities (where $n$ depends
on the length of the words in the Fibonacci language associated to the other two
systems of Ammann planes that determine the singularities), then the total region of the
IAQ considered has $N = k_j n$ threefold singularities in total. 

\smallskip

The bulk free energy determines how much work a system 
can produce and  can be computed as in \cite{Bax} and \cite{Bax3} and \cite{Kor2}.

\begin{prop}\label{bulkF} 
Given the partition function per site 
\begin{equation}\label{kappa}
    \kappa = Z^{1/N},
\end{equation}
with $N = k_j n$ as above, 
the bulk free energy for the periodic model is given by
\begin{equation}\label{f0}
    F_0 = -k_B T \log \kappa,
\end{equation} 
where $k_B$ is Boltzmann's constant and $T$ is the temperature. 
The bulk free energy $F$ of the spin system on the icosahedral quasicrystal is 
obtained as in \S 4 and \S 6 of \cite{Kor2}, in the form
\begin{equation} \label{bulk_free_energy}
F = \sum_{6\geq j>k>i\geq 1} \omega_{ikj} F_0(\theta_{1},\theta_{2},\theta_{3}),
\end{equation}
with $F_0$ as in \eqref{f0} and with
$\omega_{ikj}$ the relative frequency of appearance of a given rhombohedron, which
for the prolate and oblate rhombohedra, respectively, is given by
\begin{equation}\label{omegap}
        \omega_p = \frac{1}{10}\tau^2, \ \ \ \  \omega_o = \frac{1}{10}\tau \, .
\end{equation}
\end{prop}

\medskip

The bulk free energy $F_0$  for the periodic system is computed in \cite{Bax} and \cite{Bax3} as
\begin{equation}\label{eqFfree}
\begin{array}{ll} 
\displaystyle{ \frac{-F_0(\theta_1,\theta_2,\theta_3)}{k_BT} } =  &
  -\log(c_0c_1c_2c_3) + \Psi(\pi - s) + \Psi(s-a_1)+\Psi(s-a_2)+\Psi(s-a_3) \\[3mm] & 
  \displaystyle{ + \frac{1}{2\pi} \sum_{i=1}^{3}( a_i \log\sin(\frac{\theta_i}{2}) + (\pi - a_i)\log\cos(\frac{\theta_i}{2}) ) }\, .
  \end{array}
\end{equation}
where $\Psi(x)$ is the polylogarithm function
\begin{equation}\label{sPsi}
     \Psi(x) = \sum_{m=1}^{\infty} \frac{\sin(2mx)}{2\pi m^2}\, , 
 \end{equation}        
and let $2s= a_1 + a_2 + a_3$ is the perimeter 
of the spherical triangle defined by the angles $\theta_i$, and with
the $c_i$ as in \eqref{ctseq}. We refer the reader to \cite{Bax} and \cite{Bax3}  for more details.

\medskip

It is not expected that an explicit expression like \eqref{eqFfree} for the free energy
will have a generalization to other spin systems obtained from MCFGs as in Proposition~\ref{IAQmCFG}.
However one can consider, for a spin system of the class obtained in Proposition~\ref{IAQmCFG},
a formulation of the free energy analogous to \eqref{bulk_free_energy} in terms of 
\eqref{f0} and \eqref{kappa}.

\smallskip

In the more general setting of Proposition~\ref{IAQmCFG}, 
one can still consider the partition function per site \eqref{kappa},
with $Z$ the partition function with Boltzmann weights of the
form \eqref{BoltzWm}, and with $N$ given by the product of the
lengths of the words $\omega_i$ in a tuple $(\omega_1, \ldots, \omega_q)\in \Sigma_\cG$
describing a corresponding region of the spin model. Then one
defines a corresponding $F_0$ as in \eqref{f0}. This can be seen, for an mMCFG, as a function
$F_0(\underline{J}_{\fA^m})$, through the dependence of the Boltzmann weights \eqref{BoltzWm}
on the datum $J$, as in Proposition~\ref{IAQmCFG}.
The bulk free energy $F$ of \eqref{bulk_free_energy} will then have an analog of the form
\begin{equation}\label{bulkFmcfg}
F =\sum_{\underline{a} \in \fA^m} \, \omega_{\underline{a}} \, F_0(\underline{J}_{\fA^m}(\underline{a}))\, ,
\end{equation}
where the coefficients $\omega_{\underline{a}}$ play the role of the 
densities $\omega_{ijk}$ of \eqref{bulk_free_energy} and are given by the frequency
of occurrence of $\underline{a}$ in $m$-tuples of words in $\Sigma_\cG$.

\begin{ques}\label{ques1}
Is there an explicit bulk free energy computation for the more general setting 
of Proposition~\ref{IAQmCFG} and Theorem~\ref{CubeThm}?
\end{ques}

\medskip

\subsubsection{Criticality} \label{CritSec}
It was conjectured in \cite{BaxFor}  and shown in \cite{Bax} that the 
Zamolodchikov model is critical. Criticality for the quasiperiodic case was
shown in \cite{Kor2}, where it is observed (\S 7 of \cite{Kor2}) that the
thermodynamic functions of the quasiperiodic system have the same behavior
(but different coefficients) as those of the periodic case. 
Baxter showed criticality in \cite{Bax}
for the $n \times \infty \times \infty$ lattice, namely a lattice that is infinite in two 
spatial directions and finite in another one. 
By calculating the free energy of the $n \times \infty \times \infty$ lattice, one can prove 
the model is critical by proceeding as in \S 10 of \cite{Bax}. We do not reproduce
here the full argument, for which we refer the reader to \cite{Bax}, but the
key result is an asymptotic formula of the form
$$     \log(\kappa_n/\kappa_\infty) = \frac{2\pi (\sin\alpha_0 \sin\alpha_1 \sin\alpha_2 \sin\alpha_3)^{1/2}}{3n^2 \sin^2\theta_3} + \mathcal{O}(n^{-4}) \, , $$
with $\kappa_n$ the partition function per site for $n$-layers. Baxter then estimates 
$\log(\kappa_n/\kappa_\infty)$ in terms of the largest and next-largest eigenvalues of the transfer matrix,
respectively denoted by $\Lambda_0$ and $\Lambda_1$, showing that the relation $Z = \Tr(T^{k_j})$
implies $\log(\kappa_n/\kappa_\infty) = \mathcal{O}((\Lambda_1/\Lambda_0)^n)$, which means that if
$|\Lambda_1/\Lambda_0|<1$ then $\kappa_n$ should approach $\kappa_\infty$ exponentially fast,
while the expression above shows that it doesn't.
The argument of \cite{Bax} concludes with the observation that the large-lattice limit 
of the ratio $|\Lambda_1/\Lambda_0|$  should be strictly less than one for a non-critical model,
thus showing criticality.

\medskip

The argument for criticality used by Baxter in \cite{Bax}, that we recalled here, shows
the criticality of the spin model on the icosahedral quasicrystal. A similar question can
be formulated for other spin models associated to formal languages that are MCFGs. 

\begin{ques}\label{ques3}
Is there a version of the criticality argument recalled here for the more general 
setting of Theorem~\ref{CubeThm}?
\end{ques}

It is important, in addressing such questions, to keep in mind that
the general case of systems obtained as in Proposition~\ref{IAQmCFG} does not
correspond to actual geometries (unlike the quasiperiodic tilings of the
icosahedral quasicrystal model). In general the spin systems obtained in this way 
are only abstract models, coded by MCFGs in a similar way to how the quasiperiodic 
cases are codes by a formal language describing the geometry of their Ammann 
quasilattices but without the support of the explicit geometry.

\bigskip

\subsection*{Acknowledgment} The first author is supported by the Carl F. Braun 
Residuary Trust and the Caltech WAVE program for undergraduate research.
The second author is supported by NSF grant DMS-2104330.

\end{document}